\documentclass[aps,pre,twocolumn,amssymb, amsmath,groupedaddress]{revtex4-1}

\usepackage{natbib}
\usepackage{graphicx}
\usepackage{dcolumn}
\usepackage{amsmath,amsthm}
\usepackage{amsfonts}
\usepackage{amssymb}
\usepackage{bbold}
\usepackage{times}
\usepackage{mathrsfs}
\usepackage{color}
\usepackage{gensymb}
\usepackage{times}
\usepackage[]{graphicx}
\usepackage{bm}
\DeclareGraphicsExtensions{.PNG,.jpg,.pdf,.gif,.eps}
\usepackage{amsmath,amssymb,amsthm,amsfonts,eucal}
\usepackage[english]{babel}
\usepackage{mathtools, nccmath}
\usepackage[detect-none]{siunitx}
\def\A{{\bf A}}

\def\B{{\bf B}}

\def\Id{{\bf I}}

\newcommand{\euler}{e}

\DeclarePairedDelimiter{\nint}\lfloor\rceil

\begin{document}
%
\title{Variety of scaling behaviors in  nanocrystalline plasticity}
\author{ P. Zhang ${}^{1}$, O. U. Salman${}^{2}$, J. Weiss${}^{3}$ and L. Truskinovsky${}^4$ }
\affiliation{  ${}^1$ State Key Laboratory for Mechanical Behavior of Materials, Xi'an Jiaotong University, Xi'an, 710049, China\\ ${}^2$ CNRS, LSPM UPR3407, Paris Nord Sorbonne Universit$\acute{e}$, 93430, Villetaneuse, France\\ ${}^3$ IsTerre, CNRS/Universit$\acute{e}$ Grenoble Alpes, 38401 Grenoble, France\\ ${}^4$ PMMH, CNRS  UMR 7636, ESPCI ParisTech, 10 Rue Vauquelin, 75005, Paris, France }

\date{\today}

\begin{abstract}

 We address the  question of why  larger,  high symmetry  crystals are  mostly weak, ductile and  statistically sub-critical, while smaller  crystals with the same symmetry are strong, brittle and  super-critical.   We link it to another question of  why intermittent elasto-plastic deformation of sub-micron crystals  features highly unusual size sensitivity  of scaling exponents. 
 We use a minimal  integer-valued automaton model of crystal plasticity to show that with growing  variance of quenched disorder,  which  can serve in this case as a   proxy for increasing size,   sub-micron crystals  undergo a crossover from spin-glass  marginality to  criticality characterizing  the second order brittle-to-ductile  (BD) transition.  We argue that this crossover is behind  the non-universality of scaling exponents observed in physical and numerical experiments.  The non-universality  emerges  only if the quenched  disorder is elastically incompatible and it  disappears if the disorder is compatible.    
\end{abstract}

\maketitle

\section{Introduction}
\label{I}

Considerable research efforts  have been recently focused on the study of mechanical properties of sub-micron crystals \cite{Han2015-db,Maas2012-ib,Mordehai2011-to,papanikolaou2017avalanches, Maas2018-qu}. It was  found  that the  deformation mechanisms, which we habitually associate with dislocation plasticity, change dramatically once the sample size is reduced below  the micrometer range. Strength of such crystals was shown to be  size-dependent  \cite{Uchic2004-ax,Greer2005-ak,Dimiduk2006-fz},  with  
stress-strain response  exhibiting  pronounced  intermittency  and  
scale-invariance   over   a wide range of  scales,  independently of crystal symmetry ~\cite{Zaiser2006-gk,Uchic2009-jl,Csikor2007-jk, Friedman2012-ie,Maass2015-vl, Ng2008-ii,Brinckmann2008-od,  zaiser2008strain}.  Both measured and  computed   scaling exponents  were  shown to feature  highly unusual size dependence  \cite{Papanikolaou2012-ai,Zhang2017-cl}.  

Moreover, even though  plasticity at macroscale is generally associated with ductility, crystal  plasticity  at sub-micron  scales  exhibits major stress drops  or strain bursts reminiscent of brittle fracture ~\cite{Cui2017-xn,Maas2018-qu,wang2012pristine, chrobak2011deconfinement,bei2008effects}. Brittleness, usually attributed  to dislocation-free crystals  \cite{broberg1999cracks}, reappears in nano-particles and nano-pillars  that seem to be 'breaking  plastically'  by generating  a large number of  globally correlated dislocations  \cite{sharma2018nickel,Mordehai2018-qm}.  The implied    system-size events  hinder  our  ability to control  plastic  deformation at sub-micron scale   and compromise the  reliable  functioning of ultra-small machinery  \cite{Argon2013-fv,Csikor2007-jk,Benzerga2009-ny,Motz2009-dx,Uchic2009-jl}. 

The  peculiar properties of sub-micron crystals  can be linked to  the scarcity of dislocation sources and easiness of surface annihilation. This   limits dislocation storage and inhibits  forest hardening, thus   reducing  dislocation network complexity and  
promoting highly anisotropic single slip behavior. The lack of obstacles facilitates the  collective behavior, which is ultimately behind intermittency and scaling. 
Rationalization of the \emph{crossover}  from  bulk to surface dominated plastic flow has emerged recently as one of the  main challenges in crystal plasticity.

 To illustrate the full spectrum of   plastic responses at  sub-micron scale,   we decided to characterize them experimentally  for a \emph{single} material.   We conducted a set of compression tests    on pure Mo sub-micron pillars,   choosing intentionally  a 'mild', in the sense of \cite{Weiss2015-eh},  BCC crystal.  The main qualitative observation was  that   larger,  dislocation-rich   sub-micron crystals are  weak, ductile, and  statistically sub-critical. In contrast,   smaller, dislocation-starved crystals are strong, brittle,  and  statistically super-critical.  One of the aims of this paper is  to reproduce  the observed behavior using  a minimal, analytically transparent model.

The intermittent plastic deformation in  crystals  
was modeled  previously using  molecular dynamics \cite{Niiyama2015-gv}, discrete dislocation dynamics  \cite{Miguel2001-fi,Ispanovity2014-ra,
Ovaska2015-yb,Papanikolaou2012-ai,Cui2017-xn}, phase field theories \cite{Chan2010-ha} and   various meso-scopic  approaches  \cite{Zaiser2007-mv,Salman2011-ij}.  The   results of  different simulations are not fully consistent, suggesting that  scaling exponents  may be  covering  a  broad  range of values \cite{Sparks2018-zp,Song2019-bn,Zhang2017-cl} and supporting  the idea that   micro-plasticity is not a  universal  critical phenomenon. 

Here we show that, rather surprisingly,  the use of an oversimplified     model of crystal plasticity introduced in \cite{Salman2011-ij, Salman2012-us},  allows one  to reconcile the  existing results   while  dealing with  realistic   preparations and avoiding  ad hoc assumptions.  The main step   is the  reduction of the  plastic flow problem to a computationally  effective integer-valued \emph{discrete automaton}.  Despite the simplicity of the ensuing dynamical system,   one    can  account in this way  for both short-range and long-range  elastic  interactions, including dislocation nucleation and immobilization.   It also allows one  to     accumulate sufficient statistics,  since one can deal in this way  with  millions of meso-scopic elements and tens of thousands of dislocations.  
  
Our main result is  that  the  non-universality  of   sub-micron plasticity   and the inferred brittleness of ultra-small crystals can be conceptualized  as a  multi-stage   crossover  
 from spin-glass marginality, characteristic of very small, almost defect-free  crystals,  to the criticality of larger  crystals associated with a  brittle-to-ductile (BD)   transition.  
 
To simulate  the  size dependence of scaling exponents in small crystals we assumed  that,  at least in sub-micron range, the  decreasing  variance of the quenched disorder can  serve  as a   \emph{proxy} for contracting crystal size.    
Behind this assumption is the idea that the  role of surface annihilation of dislocations 
can be mimicked by the scarcity of external sources required for dislocation nucleation in the bulk. 
Essentially,  we exploit the fact that in small systems, the conventional dislocation nucleation sources are compromised or even disabled by their closeness to the surface.

Despite the rather prototypical  nature of  such  approach, we were able to capture both, qualitatively and quantitatively,  the  size dependence 
 observed in our experiments on Mo micro-pillars.  Rather remarkably, we   reproduced almost \emph{exactly} the measured value of the critical exponent  characterizing the BD transition in such crystals. Along the way, we revealed  a fundamental distinction between elastically compatible (local) and elastically incompatible (nonlocal) quenched disorder  and showed that the non-universality  emerges  only if the quenched  disorder is 'nonlocal'  and that it  disappears if the disorder is 'local' as, for instance, in the conventional  random field Ising model (RFIM) \cite{dahmen1996hysteresis, Ozawa2018-xi}.  
 
 It should be mentioned that some closely  related  results have  been  previously obtained    in the studies  of  amorphous glasses,  also exhibiting brittleness, yield, intermittency, and the BD transition.
 However, outside the limit of  extremely well-annealed glasses (practically unreachable), the amorphous systems remain fundamentally different  from crystals. For instance,   the quenched disorder in glasses and granular systems is  often rather special as it is revealed by the  hierarchical structure of their energy landscapes. More generally, amorphous solids  lack  long-range order,   which is behind  crystallographic  constraints for plastic slip, and which ultimately ensures orientation dependence of the mechanical response. Most importantly,  mobile dislocations,   that may  nucleate, annihilate and  form complex  entanglements in crystals, do not exist in amorphous solids \cite{Procaccia2017-gq,Ozawa2018-xi,Popovic2018-pp,Shang2020-zv}. As we show in this paper, the existence of an additional structure in crystalline plasticity, makes their scaling behavior more nuanced. 

 The rest of the paper is organized as follows. In Section \ref{FSTS} we present the results of our compression tests on Mo micro-pillars. We then formulate our computational  model in  Section \ref{M}. The possibility to use quenched disorder as a proxi for the system size is discussed in detail Section \ref{DPS}. The macroscopic stress-strain response of the system  is   studied in  Section \ref{SR} and in  Section \ref{SC} we quantify the fractal structure of the associated plastic strain fields. The disorder dependence of the statistics of avalanches is analyzed in  Section \ref{AS}.  The related distribution of stability measures is discussed in Section \ref{SD}.  A   simple mean field model,  building a bridge between  our computational results and the macroscopic parameters used in phenomenological models of crystal plasticity,  is presented in Section \ref{MF}. In Section \ref{CL} we compare the results for monotone and oscillatory loading and in Section \ref{LN} we  compare the effects on plastic  scaling  of  'local' and  'nonlocal'   quenched disorders. Our main results are summarized in Section \ref{C}.

\section{Experiment}
\label{FSTS}

\begin{figure}[!htbp]	
\includegraphics[scale=0.5  ]{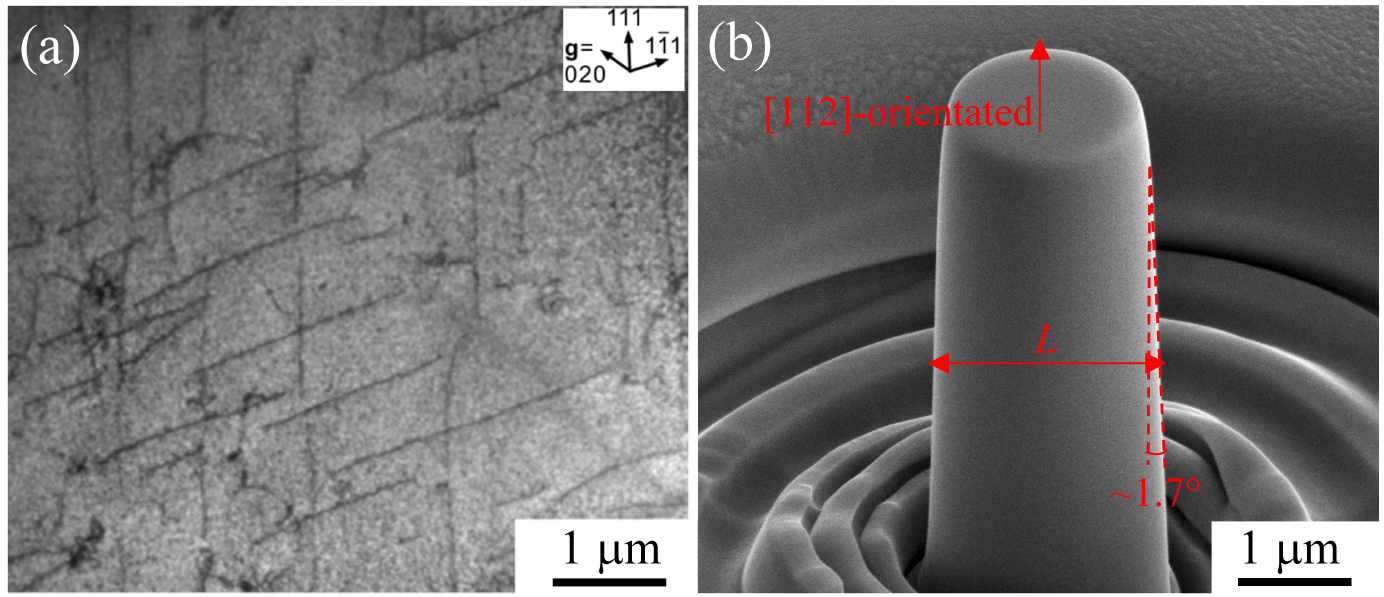}
\caption{   (a) TEM image showing the Grown-in dislocations in the bulk Mo crystal.  (b) SEM image of [112]-orientated micro-pillar before compression.  }
\label{fig1}
\end{figure}

A millimeter-sized Mo single crystal was cut from a well-annealed Mo ingot of a high purity ($>$ 99.99\%). The initial dislocation structure inside this  BCC crystal was characterized by  transmission electron microscopy (TEM), showing straight screw dislocations along $\langle 111 \rangle$ directions (Fig. \ref{fig1}(a)), with a density $\rho \approx 1.6\times10^{12} \;\rm{m}^2$ measured by the line-intercept method.  According to these data the  estimated   equidistant dislocation spacing   is  $l \sim 1/\sqrt{\rho} \approx 790 \;\rm{nm}.$ 
 
The [112]-oriented Mo pillars with diameters from 500 nm to 1500 nm were fabricated on the electropolished surface of the single bulk crystal by using a Ga-operated focused-ion beam (FIB). The height-to-diameter ratio of the pillars was kept between 2.5:1 and 3:1, and the taper was $\sim 1.7^\circ$ (Fig. \ref{fig1}(b)). A nano-indentation system (Hysitron Ti 950) was then used to compress the pillars at room temperature under controlled  displacement, with a strain rate of $2\times10^{-3} \; \rm{s}^{-1}$. 

\begin{figure}[!htbp]	
\includegraphics[scale=.6 ]{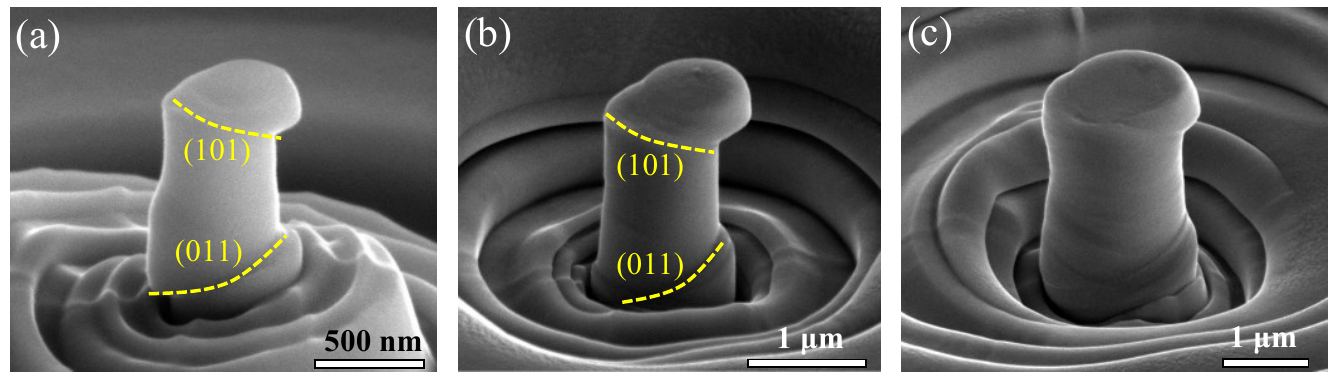}
\caption{ SEM images of [112]-orientated micro-pillars after compression: (a) 500 nm, (b) 1000 nm and (c) 1500 nm.    The marked slip traces in (a) and (b) indicate the  locally single-slip nature of the plastic flow in 500 nm and 1000 nm  crystals.}
\label{fig101}
\end{figure}

At least four samples were tested for  each value of diameter. In Fig. \ref{fig101} we show the SEM images of the  micro-pillars after   compression for each size separately. Note the highly anisotropic, single slip plane character  of the local plastic deformation pattern in 500 nm and 1000 nm samples: the  crystallographically exact slip traces are specifically indicated in Fig. \ref{fig101} (a,b)).  In 1500 nm crystals the plastic flow becomes more isotropic showing even locally  a multi-slip deformation pattern.

\begin{figure}[!htbp]
 \includegraphics[scale=0.65]{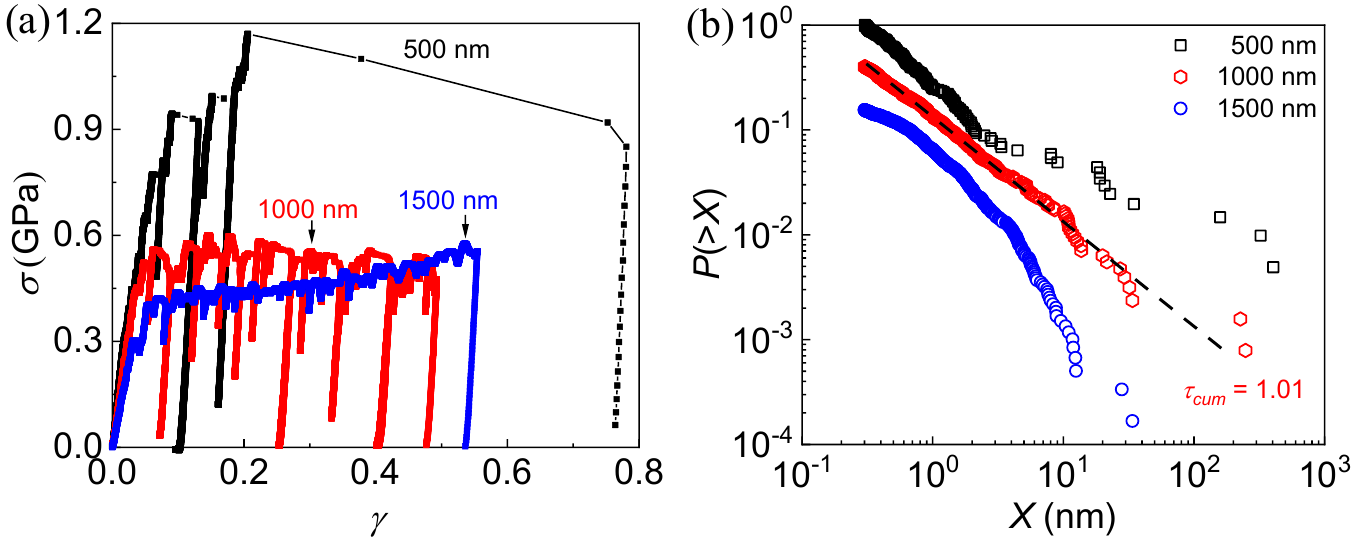}
\caption{ \label{fig1001}  Compression tests on pure Mo sub-micron pillars:  (a)    stress -  strain curves (shear); (b) cumulative  distribution of plastic displacements $ X$ detected over the entire loading process; here $\tau_{cum}=\tau_{in}-1$ where $\tau_{in}$ is the stress integrated exponent. 
}
\end{figure}

 In Fig. \ref{fig1001}(a),  we  juxtaposed  the stress-strain curves  for  Mo pillars with diameters from 500 nm to 1500 nm, all showing a characteristic set of abrupt discontinuities.  For the chosen  pillar orientation, the slip systems with maximum Schmidt factor $S$ are (101) $\langle\overline{1}11\rangle$ and (011) $\langle1\overline{1}1\rangle$. Accordingly, after deformation,  we  observed the most significant plasticity on the planes $\lbrace110\rbrace$.  

 The complex configuration of the observed jumps on the stress-strain  plane can be explained by the delayed  instrumental response  during   rapid  plastic deformation, see  \cite{Zhang2017-cl} for more details.  Here we only  briefly mention  that  the mechanical loading in such experiments is  performed  through an  auto-regulation system  with PID feedback.  The loading device  adjusts   dynamically, and in the case of an avalanche, it usually does not have enough time to respond. As a result, we observe displacement jumps at  an almost constant force.  
 
 The plastic displacement jumps were determined from the recorded force-displacement data   using the post-processing methodology developed in \cite{Zhang2017-cl}.  The size of dislocation avalanche  was associated with the plastic displacement $X=D_e-D_s+(F_s-F_e)/K_p$, where $D_s$ and $D_e$ are the measured displacement  at the beginning of the jump and at  its end,  $F_s$ and $F_e$ are the corresponding values of the  force,  and $K_p$ is the independently measured stiffness of the pillar. The computed value of $X$ is expected to  scale with the total distance covered by all mobile dislocations during an avalanche \cite{maass2013small}.

\begin{figure}[!htbp]	
\includegraphics[scale=.7]{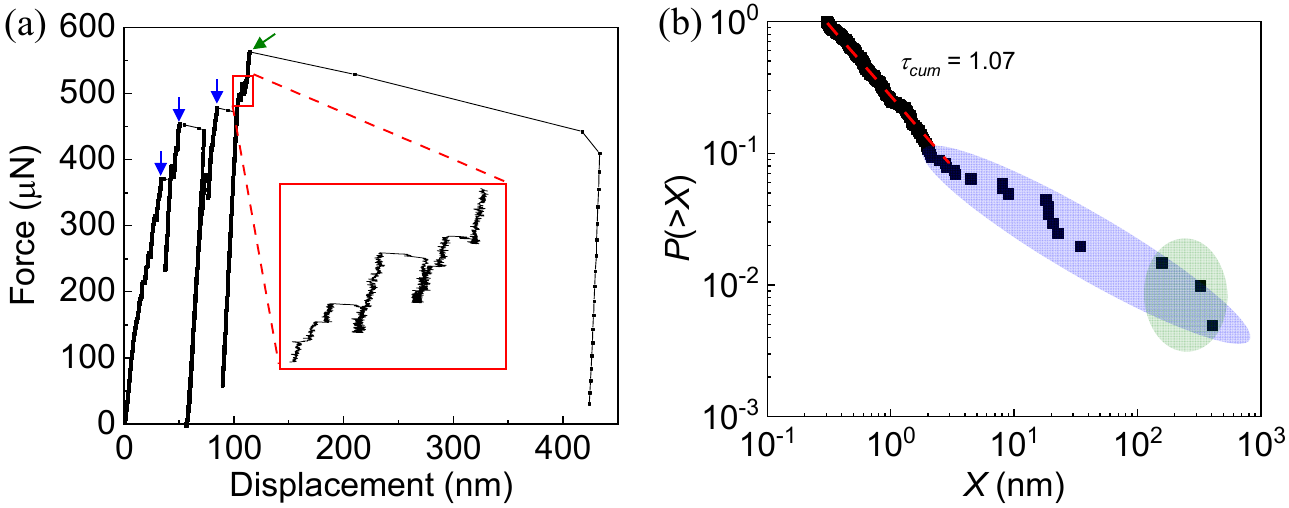}
\caption{ \label{fig202} Representative force-plastic displacement  curve  (a) and cumulative probability distribution of plastic displacements $X$ detected over the entire deformation path (b) for the Mo pillars with 500 nm diameter.  }  
\end{figure}

The cumulative probability distributions $P(X)$  are shown in Fig. \ref{fig1001}(b)  for all three sample sizes. 
A statistical analysis of these distributions, which involved  the comparison of the  p-values  and the likelihood ratios,  allows us to conclude that: (i) 
for 500 nm Mo micro-pillars  the outliers observed above $X \approx 3   \ nm$  are statistically significant  and indicate  super-criticality;  (ii) for 1000 nm Mo micro-pillars, the power-law distribution is strongly supported; (iii)  for 1500 nm Mo micro-pillars, both,  the power-law with a cut-off  and   the log-normal distribution  are more favorable than the power-law distribution, but  the log-normal distribution is more likely. 

%

In  Fig. \ref{fig202}(a),  we  illustrate the  origin of the  characteristic   peak which is typical for the super-critical response \cite{Sornette2012-kh}, see the green ellipse  in  Fig. \ref{fig202}(a) based on  the data  for  four micro-pillars with the same diameter 500 nm.  One of these large events corresponds to the system size avalanche which is marked by the green arrow in Fig. \ref{fig202} (a) and signaling the `global failure' of the sample.
 The mid-sized bursts marked in Fig. \ref{fig202}(b) in violet correspond to brittle  events indicated in Fig. \ref{fig202}(a) by violet arrows.  
 The remaining small events show at least  one   decade of  a power-law behavior with the  cumulative exponent  $\tau_{cum} \approx 1.07$. The  coexistence of a power-law range  at small scales, with a separate peak representing system size events, is typical for spinodal criticality \cite{da2020rigidity}.

We note that the statistical super-criticality of nano-scale samples was not emphasized  in the previous studies of the scaling in sub-micron crystals \cite{Zaiser2008-ov,Cui2020-gt,Sparks2019-ie}. Instead, it was  stressed that in  almost pure crystals with negligible number of dislocations, for instance, in sub-micron and nano-particles \cite{wang2012pristine,Chrobak2011-nc}, nanowires \cite{lu2011surface}, or sub-micron  pillars \cite{Bei2008-aw,issa2015situ}, the plastic deformation culminates with the formation of a system size  slip band. As we show below, both  phenomena  have the same origin and can be ultimately linked to dislocation starvation \cite{Greer2006-zt}. 

\section{Modeling}
\label{M}

To simulate the observed behavior of micro-pillars, we use the minimal model first introduced in \cite{Salman2011-ij}.

We can assume that the displacement field is  scalar because  the plastic flow of sufficiently small micro-pillars  is mainly single-slip independently of the underlying crystal symmetry and  even in the case of multi-slip orientation. Due to a limited number of available dislocation sources within the confined volume,  the first activated slip  plane dominates and prevents other slip planes from getting involved. In this situation,  the usual  frustration leading to hardening, can be avoided considering the absence of dislocation cross-slip and facile annihilation at a free surface.  While any adequate crystal plasticity model would effectively reduce to our constrained single-slip theory in a  sufficiently small system, it should, of course,  allow for multi-slip flow  to take over at larger  sample sizes.

We assume that the crystal is oriented for a single slip along the  only available slip direction. It is modeled as an $N\times N$ square  lattice with  the meso-scopic  spacing  normalized to unity.  The deformation of the crystal is given by the displacements of the vertices of the mesoscopic elements, $\vec{u}_{i,j}=(u^x_{i,j},u^y_{i,j})$, where $i,j=1,2,\dots,N$. 

In view of our single slip assumption we  only allow displacements in the horizontal direction by setting $u^y_{i,j}\equiv0$. We can then introduce the notation  $u_{i,j} \equiv u^x_{i,j}$. In the presence of a  kinematic constraint the strain tensor can be reduced to  two fields: a longitudinal  strain, 
 $\zeta_{i,j} =
u_{i+1,j}-u_{i,j},$  
which is a  linear, non-order parameter variable,  and a shear strain 
 $\xi_{i,j} =u_{i,j+1}-u_{i,j},$ 
which is a nonlinear,  order parameter type variable,  given that  plastic slip originates from multi-well  nature of lattice potential. 

We write  the dimensionless  energy of the system in the form \cite{Salman2011-ij}
\begin{equation}
\label{eq:PHI}
\Phi=\sum_{i,j} f(\zeta_{i,j},\xi_{i,j}),
\end{equation}
where 
 \begin{equation}
f(\zeta,\xi)=( K/2)\zeta^2+f_0(\xi)
 \end{equation}
is the energy of a single (meso-scopic) element. To account for the lattice periodicity we assume that  $f_0(\xi)=f_0(\xi+n),$ where $n \in \mathbb{Z}$ is an integer-valued   slip. Moreover, for analytical transparency we assume that 
the \emph{periodic} energy density $f_0$ is   piece-wise quadratic  
 \begin{equation}
f_0 (\xi_{i,j}) =   (1/2)(\xi_{i,j}-d_{i,j}(\xi))^2, 
 \end{equation}
Here the plastic slip $d$   is represented  by an  integer nearest to $\xi$ so that  
$
   d_{i,j}(\xi)=\nint{\xi_{i,j}}. 
$
  
    The obtained model depends on a  single  dimensionless    parameter   $K$which  mimics the  ratios of elastic constants $ (C_{11}-C_{12})/(4C_{44})$ or $C_{11}/C_{66}$.  It  describes the  coupling between  mesoscopic elements that carry different  values of   $\xi$. In the  limits  $K \to 0,\infty$  we obtain  solvable 1D   models with mean field type interaction \cite{Puglisi2005-lg, Salman2012-us}.  At $K \neq 0$  the  model reproduces Eshelby-type propagator and therefore captures crucial effects of long range interactions   induced by elastic compatibility, see more about this in Section \ref{LN}. In our numerical experiments we assumed that $K=  2$ which represents  a typical value for  metallic crystals.

\begin{figure}[!htbp]	
	\centering
	\includegraphics[scale=0.3]{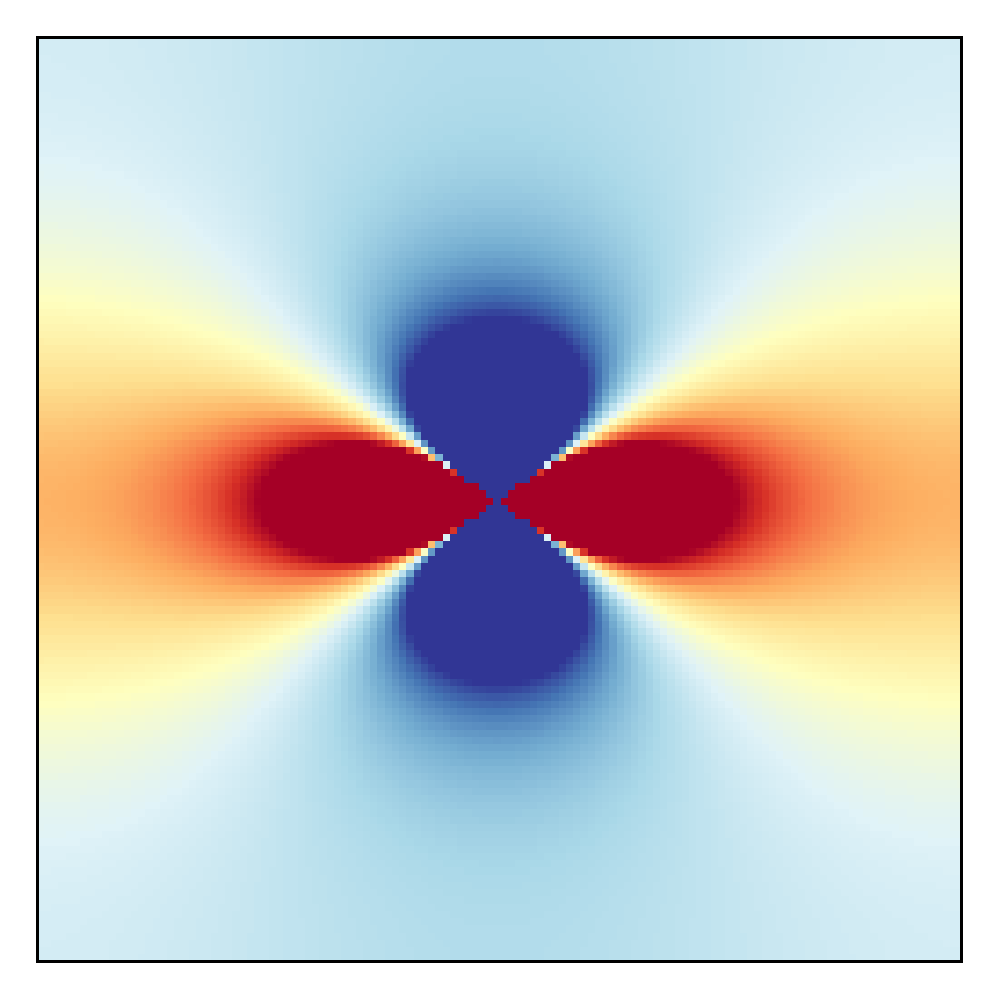}
	\caption{The real-space representation of the dipole Fourier kernel $\hat L (\bold q)$. Coloring: blue - negative, red - positive.}
		\label{kernel}
\end{figure}

The model can be reduced to a discrete  automaton  because  the elastic   problem  \begin{equation}
\partial \Phi/\partial u_{i,j}=0
\label{equil}
 \end{equation} 
 can be solved analytically  if the  integer-valued field $d$ is known \cite{Salman2011-ij}.  The associated equilibrium equations in the bulk, written terms of the displacement field $u_{i,j}$, read  
\begin{multline}
 K {}(u_{i+1,j} + u_{i-1,j} - 2u_{i,j}) + (u_{i,j+1} + u_{i,j-1} 
- 2u_{i,j})  \\
  - (d_{i,j} - d_{i,j-1}) =0.
\end{multline}
The whole system can  be  written in  matrix form 
$\textbf{M} u =b,
$
where $\textbf{M}$ is a pentadiagonal matrix and $b$ is a vector of size $N\times N$ incorporating  the boundary conditions and the field $d$. The problem then reduces to a simple matrix inversion.

We assume periodic boundary conditions  in the horizontal direction $u_{1,j} = u_{N+1,j}$.  The hard device type loading will be applied through the boundary condition in the vertical direction $u_{i,N+1} = u_{i,1}+\gamma$, where $\gamma$ is the control parameter. Periodicity is assumed to allow for the fully explicit  inversion of the matrix $\textbf{M}$. Indeed, we can then use the spectral approach based on  the Fourier transform      \begin{equation}
\hat x(\bold q)= N^{-2}\sum_{ab}x_{a,b}\euler^{-{i} \bold q \bold r}
 \end{equation}
  with $\bold  r=(a,b)$ and $\bold q=(2\pi k/N,2\pi l/N)$.  In Fourier space the solution of our linear problem is straightforward and we  can   obtain an explicit representation for   the equilibrium  shear strain
 \begin{equation}
 \hat\xi(\bold q) = \gamma\delta(\bold q) + \hat L (\bold q) \hat d(\bold q), 
 \end{equation}
 where we recall that  $\gamma=\langle\xi\rangle$  is  the measure of the imposed affine deformation. The  sign-indefinite  Eshelby-type   kernel with $r^{-2}$  far field asymptotics 
 \begin{equation}
\hat L (\bold q)  = \frac{\sin^2(q_y/2)}{K\sin^2(q_x/2) + \sin^2(q_y/2)},
\label{kernel1}
 \end{equation}
is    illustrated  in the physical space  in Fig.\ref{kernel}.  Its dipolar structure  reflects the \emph{scalar} nature of our model; the more conventional  quadruple structure of the stress propagator is a feature of isotropic  elasticity, while  here we  deal with extremely anisotropic limit
\cite{picard2004elastic,tyukodi2016depinning}.  

\begin{figure}[!htbp]	
\includegraphics[scale=0.22]{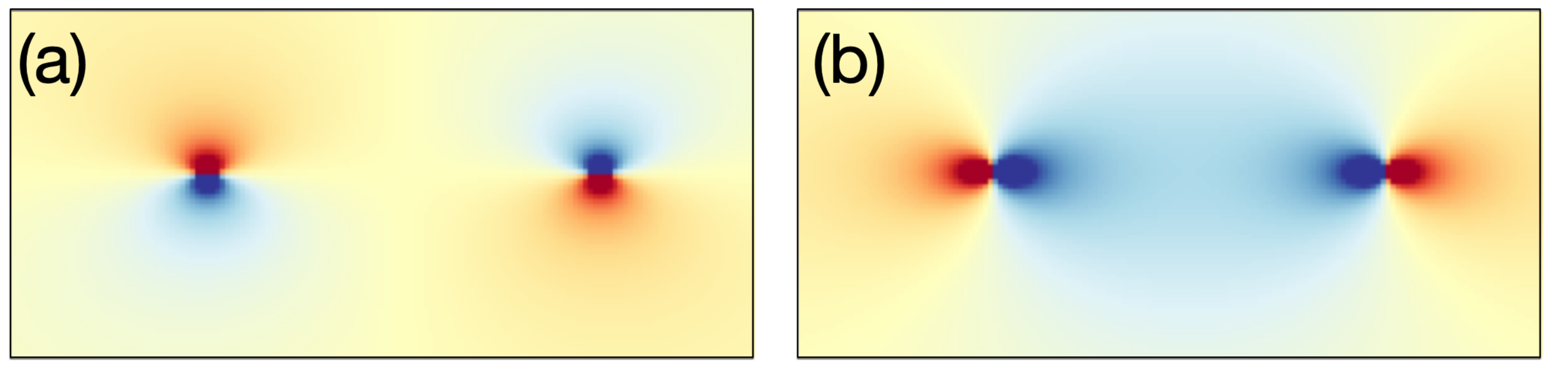}
\caption{Stress field produced by a pair of dislocations of different signs    created using a  distribution of plastic slip ($d$ field)  which vanishes everywhere except in the  one element-wide  80 lattice spacings long segment.     (a)   Axial stress  field $\sigma_{xx}$;  (b) Shear stress field  $\sigma_{xy}$.    Coloring indicates the level of  stress: blue - negative, red - positive. \label{fig6}}
\end{figure}

To illustrate dislocation nucleation in this model we show in Fig. \ref{fig6}  two dislocations of opposite signs forming a 2D topologically neutral 'loop'. 
 The  far-field asymptotics around each of the dislocations agrees with the  classical continuum  prediction $r^{-1}$, while  inside the  cores, located around the units  where  $d_{i+1,j}-d_{i,j}\neq0$,  the stress  remains finite due to the strongly discrete nature of our theory.
 
%

Since we know  how to update the elastic fields, we can  formulate the quasi-static  athermal dynamics in the form of a discrete \emph{automaton} for the integer-valued field $d$. We 
 start with the unloaded ($\gamma=0$) and   dislocation-free  state ($d_{i,j} \equiv 0$). We then advance the loading parameter $\gamma$ and compute (predict) the elastic field $u_{i,j}$ while keeping the field $d_{i,j}$ fixed. 
 The knowledge of the shear strain field $\xi_{i,j}$ allows us to update (correct) the plastic strain field using the relation $d=\nint\xi$; the update takes place when  the boundary of the energy well is reached   by at least one of the mesoscopic elements.  Then an avalanche occurs while  we use synchronous dynamics for the updates of  $d_{i,j}$. We repeat the prediction-correction steps at a given $\gamma$ till the corrections stop changing the field $d_{i,j}$ and the system stabilizes in a new equilibrium state.  As  the stress in this state  is globally below the threshold, we  can start a new search for the  increment of  $\delta\gamma$ that   destabilizes at least  one unit. As soon as 
such an element with  $d_{i,j} \neq \nint\xi$ is obtained we apply  our  relaxation protocol again,  initiating another avalanche.  When avalanche finishes, the variation of $\gamma$ resumes.

\section{Incompatible disorder}
\label{DPS}

Given our  periodic boundary conditions, we effectively consider an infinite
crystal and our parameter $N$ cannot be associated with the crystal size $L$. To model the physical size effect,  we would have to  deal with more complex   boundary conditions compatible with, say,   surface dislocation nucleation and  the formation of one-legged Frank-Reed loops.  Without such major modifications of the model,  we can study the size effect only \emph{indirectly} and we propose to  use the strength of quenched disorder as a way to differentiate between sub-micron crystal sizes.

To justify this approach we first note  that instead of $L$ one should use   a  dimensionless parameter which we can always  write as 
  \begin{equation}
R= L/l, 
\end{equation}
 where $l$ is some appropriately chosen internal length scale and without loss of generality we can write   $l \sim Gb/\sigma_{th}$,  where $G$ is the shear modulus, $b$ is  the   Burgers vector, and   $\sigma_{th}$ is the  internal stress threshold. 
 
 Following \citep{Zhang2017-cl},  we identify this  threshold with the  pinning (immobilization)  stress.  The  distinctly brittle regime for semi-pure crystal would then correspond to  small $\sigma_{th}$ due to negligible number of obstacles ensuring that $R  \ll1$.  Given that   in our Mo samples the spacing of  immobile dislocations $ l \sim 790$ nm,  brittleness of  500 nm pillars would be  in a basic agreement with such criterion. In the strongly ductile regime we should have  $R \gg 1$,  which is close to being the case for our   1500 nm   pillars.   Dislocation interaction with obstacles becomes relevant  when   $R \sim  1$  which is  the case for our  1000 nm   pillars, see  Fig. \ref{fig1}.


The threshold $\sigma_{th}$ naturally  depends on the presence of the pinning obstacles and, in general  \cite{Zhang2017-cl},  increases with  the variance of  quenched disorder imitating such obstacles. More specifically,  the decrease of  $\sigma_{th}$  can be achieved  by making the disorder more narrow which can be viewed as the way to eliminate  particularly strong obstacles. In this way, 
instead of decreasing  $L$ we  can  increase  $l$,  which should  be as effective in moving from the brittle  regime, where  $R  \ll1$, to the ductile regime,  where $R \gg 1$. In other words,  instead of exploring directly the dominance of surface effects  one can   exploit  the indirect effect  that  in smaller systems there are fewer strong obstacles that can serve, for instance,  as dislocation nucleation sites because the  existing ones are   compromised or even disabled by their closeness to the surfaces.
%
%
%
%
%
%
%

It  has to be mentioned, however,  that our association of the variance of disorder with crystal size is exclusively targeting systems without bulk criticality, as in the case of Mo crystals.  One can, in principle, manufacture  small crystals with strong (dense) quenched disorder \cite{Zhang2017-cl} or grow almost pure large crystals with very weak (sparse) quenched disorder \cite{ weiss2019ice}.  In general, both quenched disorder and  the crystal size  would  affect  brittleness, even though to grow almost defect free crystals (without solutes, precipitates and  dislocations),  is almost impossible except in case of extremely  small sizes (nano-particles). 

%
%
%


Our  crucial assumption  allows one  to study the size effect without actually changing the size of the computational system while varying instead the strength of the quenched disorder. Here by disorder we mean first of all   inclusions such as  solute atoms. 

It  can be also impurity atoms or  even  vacancies (or voids) resulting from the motions of dislocation jogs \cite{deschamps2012situ}.  It will be  important, however,  that   such disorder is elastically incompatible producing long-range effects as, for instance,  in the case  of  local volumetric changes.

To account for such  disorder in the most simple linear form, we can add to the energy density  an  additional   term proportional to the non-order parameter variable $\zeta$ obtaining
 \begin{equation}
 f (\xi_{i,j},\zeta_{i,j} ) = \frac{K}{2}\zeta_{i,j}^2+\frac{1}{2}(\xi_{i,j}-d_{i,j}(\xi))^2-h_{i,j}\zeta_{i,j}.
 \end{equation}
Here the   random coefficients  $h_{i,j}$,   
drawn independently in each lattice point from  Gaussian distribution with variance $\delta$   
\begin{equation}
p(r)= (2\pi\delta^2)^{-1/2}\exp{(- r^2/(2\delta^2))},
 \end{equation} 
 mimic  incompatible  lattice pre-stress  which  acts  on the order parameter variable $\xi$  only \emph{indirectly}.   
 
 \begin{figure*}[!htbp]
\includegraphics[scale=0.55]{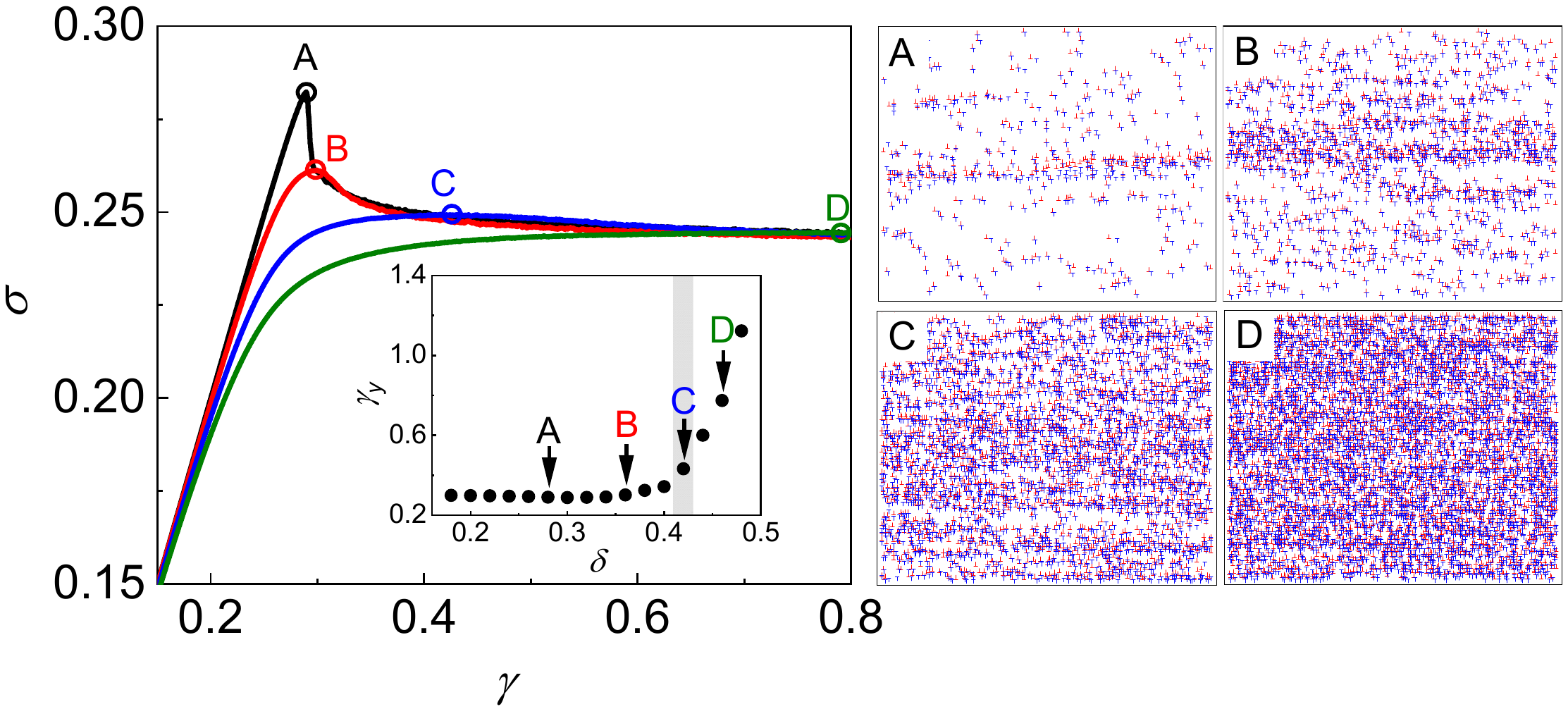}
\caption{The effect of disorder on the average strain-stress curves in simple shear  tests. Inset  shows the   yield strain  $\gamma_y$  (points A, B, C, D); the gray strip schematically marks the BD transition. The  zoom on the  'after yield'  dislocation configurations   is shown in the right column where   the total number of dislocations  is  782 (in A), 2600 (in B), 5220 (in C) and 8862 (in D). The dislocations with positive and negative Burgers vector are marked by red and blue. Parameters: $K=2, N=1024$.  \label{fig2}}
\end{figure*}

In the presence of quenched disorder $h$,  the strain fields take the form
\begin{equation} 
 \hat\xi(\bold q)=\gamma\delta(\bold q)+\frac{\hat s_y^+(\bold q)[\hat s_x^-(\bold q)\hat h(\bold q)+\hat s_y^-(\bold q)\hat d\bold(\bold q) ]}{2K[\cos(q_x) -1] +2[\cos(q_y) -1]},
 \label{xi1}
 \end{equation}
\begin{equation}
\hat\zeta(\bold q)=\frac{\hat s_x^+(\bold q)[\hat s_x^-(\bold q)\hat h(\bold q)+\hat s_y^-(\bold q)\hat d\bold(\bold q)] }{2K[\cos(q_x) -1] +2[\cos(q_y) -1] },
 \end{equation}
where 
   \begin{equation}
   \hat s_a^\mp(\bold q)=\pm[1-\cos(q_a)\pm i \sin(q_a)] 
 \end{equation} for  $a=x,y$. 
%
%
We can also rewrite \eqref{xi1} as
\begin{equation} 
 \hat\xi(\bold q) = \gamma\delta(\bold q) + \hat L (\bold q) \hat d(\bold q)   + \hat L_h (\bold q)  \hat h(\bold q), 
\label{automaton3}\end{equation}
where 
\begin{equation} 
\hat L_h (\bold q)  = \frac{\sin(q_x/2)\sin(q_y/2)(\cos(\frac{q_x-q_y}{2})-i\sin(\frac{q_x-q_y}{2}))}{K\sin^2(q_x/2) + \sin^2(q_y/2)} 
\label{fk2}
\end{equation} 
is a distorted  Eshelby  propagator  \eqref {kernel1} maintaining, however, its sign-indefiniteness and the decay rate  $1/r^2$.

\section{Stress-strain response}
\label{SR}

Starting with a  dislocation-free  crystal   ($d\equiv 0$) we  now drive the system quasistatically  using the  athermal quasi-static protocol  described above \cite{Puglisi2005-lg, maloney2006amorphous}.  
   The obtained  results  are then  averaged over  $\numrange{100}{3000}$ realizations of  the quenched  disorder. 
 
 In Fig.~\ref{fig2}, we illustrate  the  average stress-strain relations 
$
\sigma(\gamma)=d\Phi/d\gamma,
$
 where the   stress was averaged over the strain interval $\sim 10^{-4}$.  At each value of disorder strength $\delta$ the  stress-strain curve exhibits a maximum which we conditionally identify with the yield point. Four of such points are shown in Fig.~\ref{fig2}: $A,B,C,D$. The corresponding yield strain is denoted by $\gamma_y$ and its dependence  on disorder is shown in the inset where the states $A,B,C,D$ are also indicated.  The  zoom on the  'after yield'  dislocation configurations  in these states  is shown in the right column. 


For weak disorder, $\delta \leq 0.3$,  mimicking small, almost pure crystals,
 yielding is abrupt and brittle, accompanied by a macroscopic stress drop at the yield point and robust strain localization within a formation of a shear band (regime $A$).  With increased disorder (regime $B$),  the  first order phase transition  eventually terminates at a   critical point located around $\delta \sim 0.42$ (regime $C$), see \cite{Ozawa2018-xi,Popovic2018-pp} for  similar behavior   in amorphous plasticity. At  even stronger disorder, $\delta \geq  0.5$, representing bulk  samples,  yielding is  gradual and plasticity is ductile  with slip  uniformly distributed  over the whole crystal (regime $D$).


%

\begin{figure*}[!htbp]	
	\centering
	\includegraphics[width=13. cm]{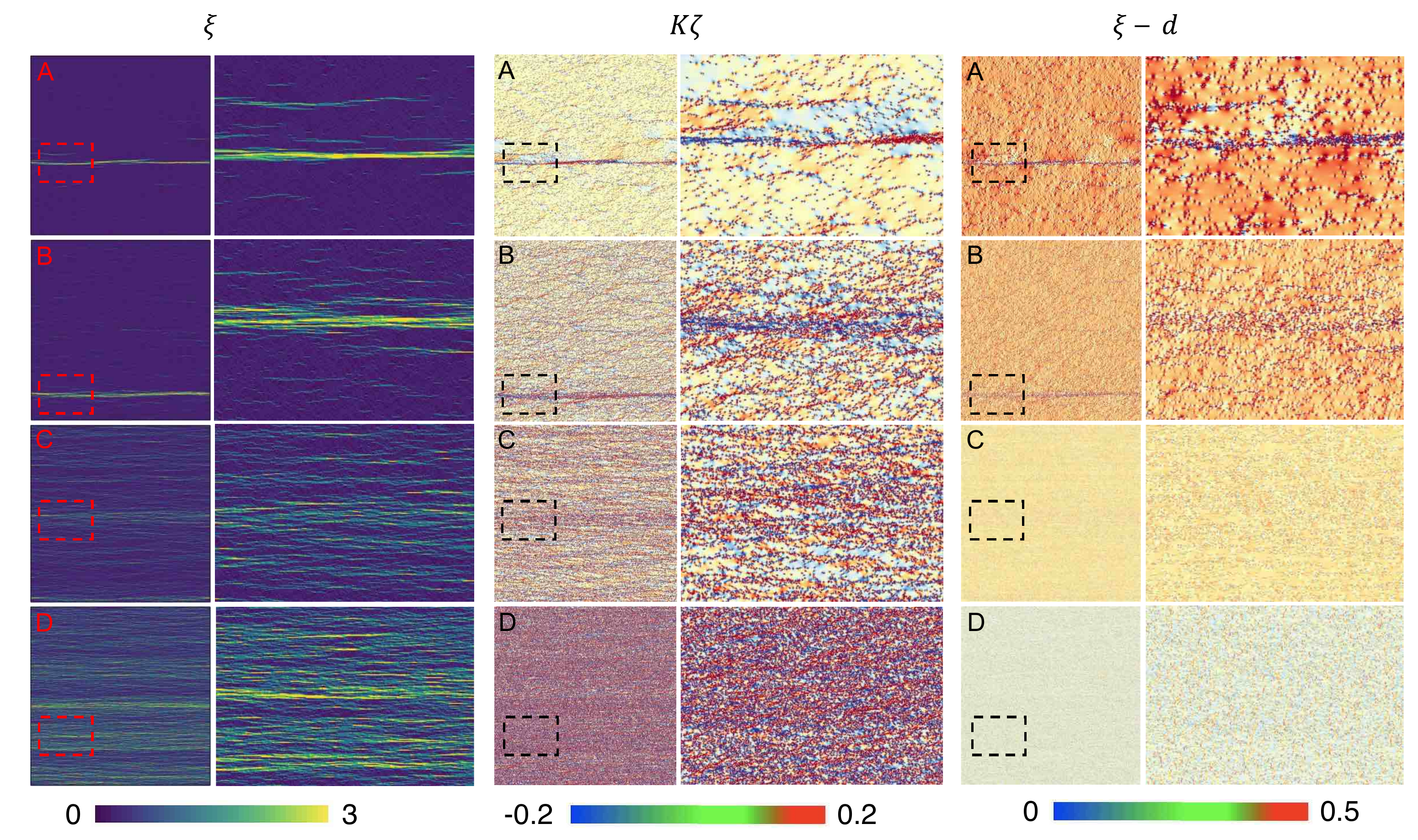}
	\caption{After yield spatial patterns: shear strain $\xi$,  longitudinal stress $\sigma_{xx}=  \text{d}f/\text{d}\zeta$ and shear stress $\sigma_{xy}=  \text{d}f/\text{d}\xi$ in the reference coordinates. Strength of disorder: $\delta=0.28$ (A), $\delta=0.36$ (B), $\delta=0.42$ (C) and $\delta=0.46$ (D).  Zoomed views of the marked areas are shown in parallel  columns. }
		\label{fig44}
\end{figure*}

In   Fig.~\ref{fig44},  we show the  shear strain, axial stress field  and   shear stress  patterns  for samples with disorder strength  marked  by the letters $A, B, C ,D$.  In these images, the affine component of the fields was subtracted. We also present   enlargement of the  marked out windows.  

When the disorder is weak (regime $A$), the shear band features a crack-like arrangement of dislocations. Outside the band, the dislocations distribution is relatively uniform, although one can trace few incipient shear pre-bands. As the strength of the disorder increases (regime $B$), the dislocation density inside the shear band diminishes and it becomes progressively broader. One can interpret this broadening as an outside  propagation of the shear band boundaries.  In  the  regime $C$, we lose a singular band which is  replaced by a diffuse network of  interconnected   pre-bands which fill the whole domain. Finally, in the ductile phase, regime $D$,  no coherent pattern is apparent as  we see dislocation activity all over the domain.  Note  that the  overall delocalization of the plastic flow, which we observed experimentally  while  increasing $L$,  is recovered here as we   increase $\delta$.
 
 
 Note, however, that the agreement between this oversimplified  theory and the experiment  cannot be complete. For instance, in our physical experiments with  sub-micron pillars, we observed  repeated almost-yielding events. During such events 
  dislocations could always  annihilate on free surfaces, which was bringing  the crystal  into the dislocation starvation state over and over again \cite{wang2012pristine}.   Instead,  in our computer experiments, where we used periodic boundary conditions, such resetting  did  not happen because 
   the crystal  could form   system-size slip bands with high dislocation density.  Therefore, for each realization of disorder  instead of several  large bursts,  we observed  a single catastrophic one.


\section{Spatial complexity}
\label{SC}

As we see plastic flow proceeds through incessant mechanical destabilization and re-accomodation of  dislocational microstructures. Due to the presence of long-range elastic interactions,  these microstructures are not random  and to reveal the nature  of  the implicit  correlations, we performed a multi-fractal analysis of the field $d_{i,j}$. 
Originally developed in the studies of fluid turbulence, 
such analysis  has become   a powerful tool   of quantifying  the degree of clustering \cite{paladin1987anomalous, mandelbrot1989multifractal, meneveau1991multifractal, lopes2009fractal}.
	
Multifractals were introduced  to study  the distribution of a scalar quantity represented by a measure 
(usually density, but in our case,  plastic strain field). The effect to capture is  that local singularities of different strengths  are distributed on  sets with  different fractal dimensions denoted by  $D_q$.
The first of those dimensions $D_0$ is the fractal dimension of the geometrical support of the measure.  As $q$ increases, the  dimensions $D_q$ become more  and more controlled by the most densely filled domains. An increasing difference between $D_0$ and $D_q$ with $q>0$ reveals  an increasingly  multi-scale   nature of the distribution.

 To compute the   dimensions $D_q$ in our case  we need to cover the deformed lattice with a regular array of boxes of size $L_b$  and sum plastic strain  in the $m^{th}$ box to obtain  $\mathcal{M}_m(L_b)=\sum d$, where the sum is taken over the  mesoscopic units covered by  a given  box. Then, the density of  plastic strain  associated with the   $m^{th}$ box is 
\begin{equation}
p_{m}(L_{b})=\frac{\mathcal{M}_m(L_{b})}{\sum_{k=1}^{n\left(L_{b}\right)} \mathcal{M}_k(L_b)},
\end{equation}
where $n(L_b)$ is the number of boxes covering the lattice. The moments of order $q$ of this density distribution  are 
\begin{equation}
M_{q}(L_{b})=\sum_{m=1}^{n(L_{b})} p_{m}^{q}(L_{b}).
\end{equation}
If the deformation pattern is self-similar, we should observe the scaling  
\begin{equation}
M_{q}(L_{b}) \sim L_{b}^{(q-1) D_{q}},
\end{equation}
 which defines the  dimensions $D_q$.
The singular value  $D_1$ can be  defined as the proportionality coefficient between $\sum_{m}^{n(L_b)}{p_m\log(p_m)}$ and $\log(L_b)$, and then  $D_{q}  \rightarrow D_{1}$ as $q \rightarrow 1$ ~\cite{Weiss2001-ha,Lebyodkin2006-ab}.  

\begin{figure}[!htbp]	
	\centering
	\includegraphics[width=8. cm]{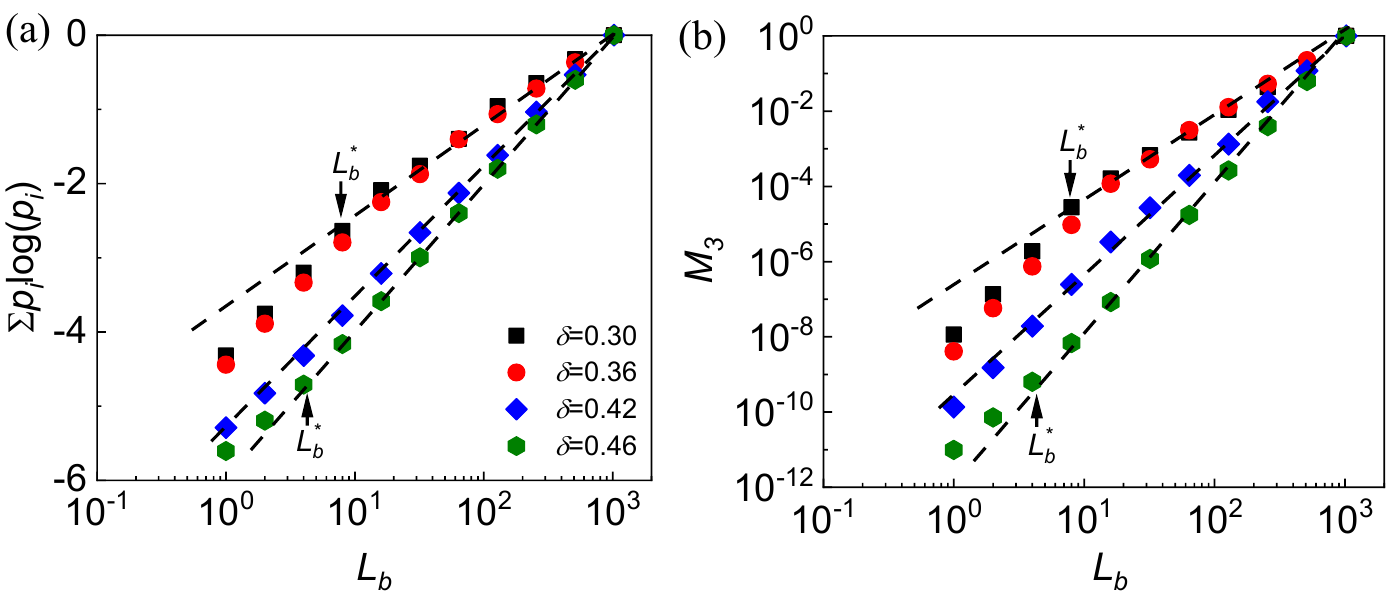}
	\caption{ Multifractal analysis of plastic strain patterns for the strain $\gamma=0.8$. (a) $\sum{p_i\log(p_i)}$ (``$M_1$'') and (b) $M_3$ as a function of box size $L_b$.}
		\label{fig5}
\end{figure}

We computed the dimensions $D_q$ at  a particular  value of the loading parameter  $\gamma=0.8$,  where  the steady flow conditions have  already been  achieved for all representative values of disorder $\delta$. Our results,  summarized  in Fig.~\ref{fig5}, clearly show the anticipated  scaling along several decades  till the cut-off scale $L_b^\ast(\delta)$. It  characterizes the  spacing of the   slip traces,  when disorder is weak,  and  the spacing of the  strong dislocation locks, when disorder is strong.  The  absence of  the cut-off  at  $\delta=0.42$ suggests  the emergence of  a scale-free hierarchical micro-structure.  

\begin{figure}[!htbp]
\includegraphics[scale=0.8 ]{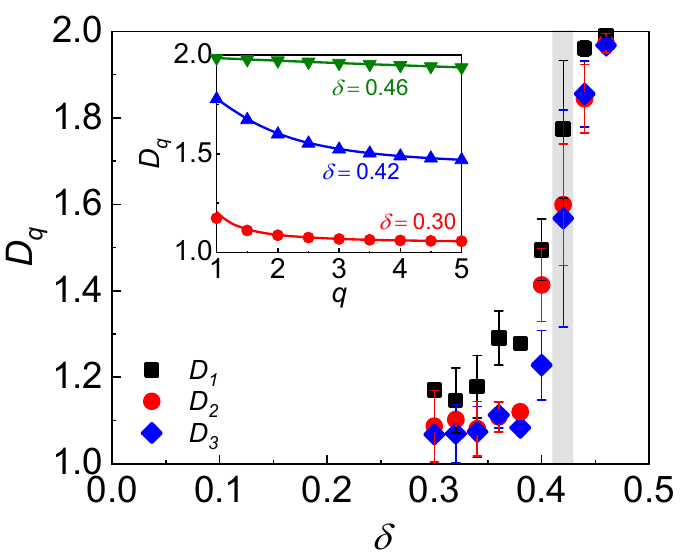}
\caption{ \label{fig41}    Disorder dependence of the fractal measures $D_q$   at $\gamma= 0.8$; inset reveals multi-fractality at  critical disorder. The gray strip schematically marks the BD transition.}
\end{figure}

In Fig.~\ref{fig41},  we show the disorder dependence of the  fractal dimensions  $D_q$.
For weak disorder  $D_q \sim 1$ for all  $q\geq1$   which signals an extreme localization. At the BD transition  the  functions $D_q$   jump towards the value  $\sim2.0$, which  indicates that the strain  pattern becomes  homogeneous. With the narrow range of disorder strengths $\delta \sim 0.42$ we can  associate   the  emergence of a turbulence-type multi-fractal pattern with $D_1>D_2>D_3>...$.   Regions of maximum  plastic strain  spatially cluster on a set with fractal dimension  $ \sim 3/2$, which is also characteristic  of  some other scale-free  systems 
\cite{salerno2013effect,gimbert2013crossover}.

\section{Avalanche statistics}
\label{AS}


Both  physical and  numerical experiments  reveal  that  quasistatically driven crystals  deform  intermittently via  avalanches  reflecting  destruction and rebuilding of  dislocation structures. To perform  a quantitative  comparison of the two types of experiment we  use our observation that  in the automaton model the energy $E$, released during an avalanche,   scales with the cumulative distance covered by  the concurrently moving dislocations.  The  distribution $p(E)$ computed in numerical experiment will then  be the analog of the experimentally measured  distribution $p(X)$.

\begin{figure}[!htbp]	
\includegraphics[width=9cm]{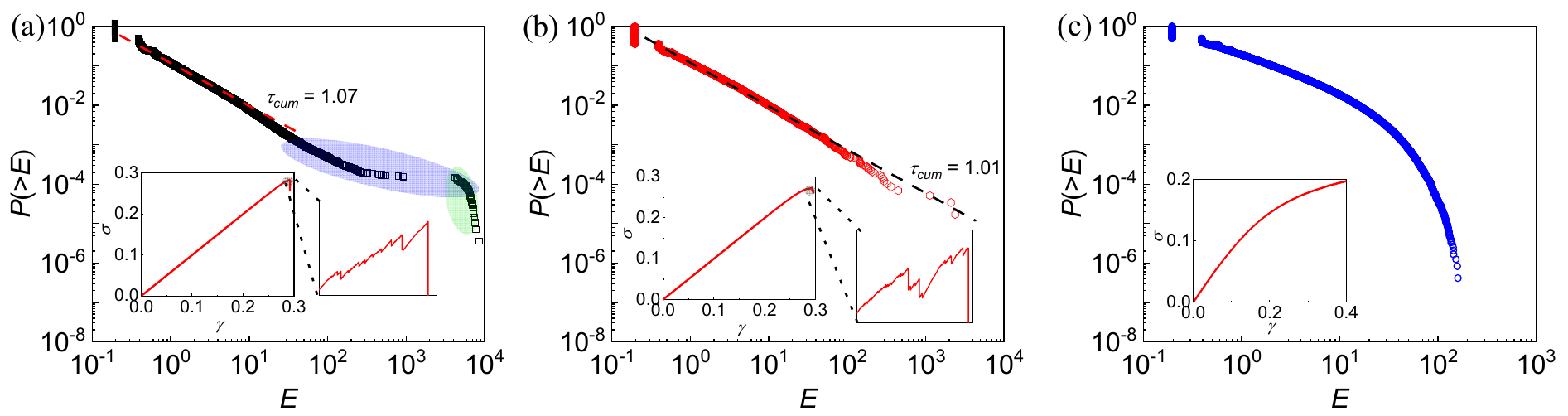}
\caption{  \label{fig_model_distribution}  Cumulative probability distributions of pre-yield avalanches  at  $\delta=0.28$(a), $\delta=0.32$(b), $\delta=0.7$(c); to be compared with the data for 500 nm, 1000 nm and 1500 nm Mo samples shown in Fig.~\ref{fig1001}(b). Averaging was performed over  100 realizations of disorder;  insets show the  stress strain curves for a particular realization of disorder. 
} 
\end{figure}
In Fig.~\ref{fig_model_distribution} we present  the three main types of  cumulative distributions  $P(>E)$ that emerged   in our numerical experiments. Comparison with Fig.~\ref{fig1001}(b) shows that they reproduce rather faithfully  all  three  types of the  distributions $P(>X)$ recorded  experimentally. Thus, at small disorder strength,  our computational  model captures the experimentally observed coexistence of characteristic bursts (SNAP events) with power-law distributed small avalanches observed in  500 nm crystals and, moreover,   predicts  a  realistic value of the experimentally measured exponent, see Fig.~\ref{fig_model_distribution}(a). With increasing disorder the numerically obtained distribution  acquires a   power-law structure, see  Fig.~\ref{fig_model_distribution}(b), with  the same exponent as in our  data obtained from  1000 nm samples, see  Fig.~\ref{fig1001}(b). At even  larger strength of disorder we observe in our numerics the emergence of subcritical  statistics, see  Fig.~\ref{fig_model_distribution}(c). Therefore,  the model   captures the observed behavior of 1500 nm samples dominated by  largely uncorrelated  POP events, see Fig.~\ref{fig1001}(b). The obtained agreement suggests that we are dealing here with very robust features of the system that are immune  to structural details and indifferent to numerical values of parameters. Our results also  strongly suggest that the size effect in micro-pillars can be indeed successfully modeled by varying the strength of quenched disorder.

We   now turn to the study of the fine details of the disorder-induced crossover phenomena  that are  not readily accessible in physical experiments. To this end, we approximate   at each level of disorder the computed stress-resolved distributions for the  released energy $E$   by the  scaling relations  $p(E)\sim  E^{-\tau}\exp(-E/E_c)$  and extract  the  disorder-dependent exponents $\tau$ using  the  maximum likelihood method \cite{clauset2009power}. A clear  advantage of the numerical experiment, where we could generate at least $3 \times 10^6$ events for each value of the exponent,  is the quality of the statistics.

\begin{figure}[!htbp]	
\includegraphics[scale=0.45]{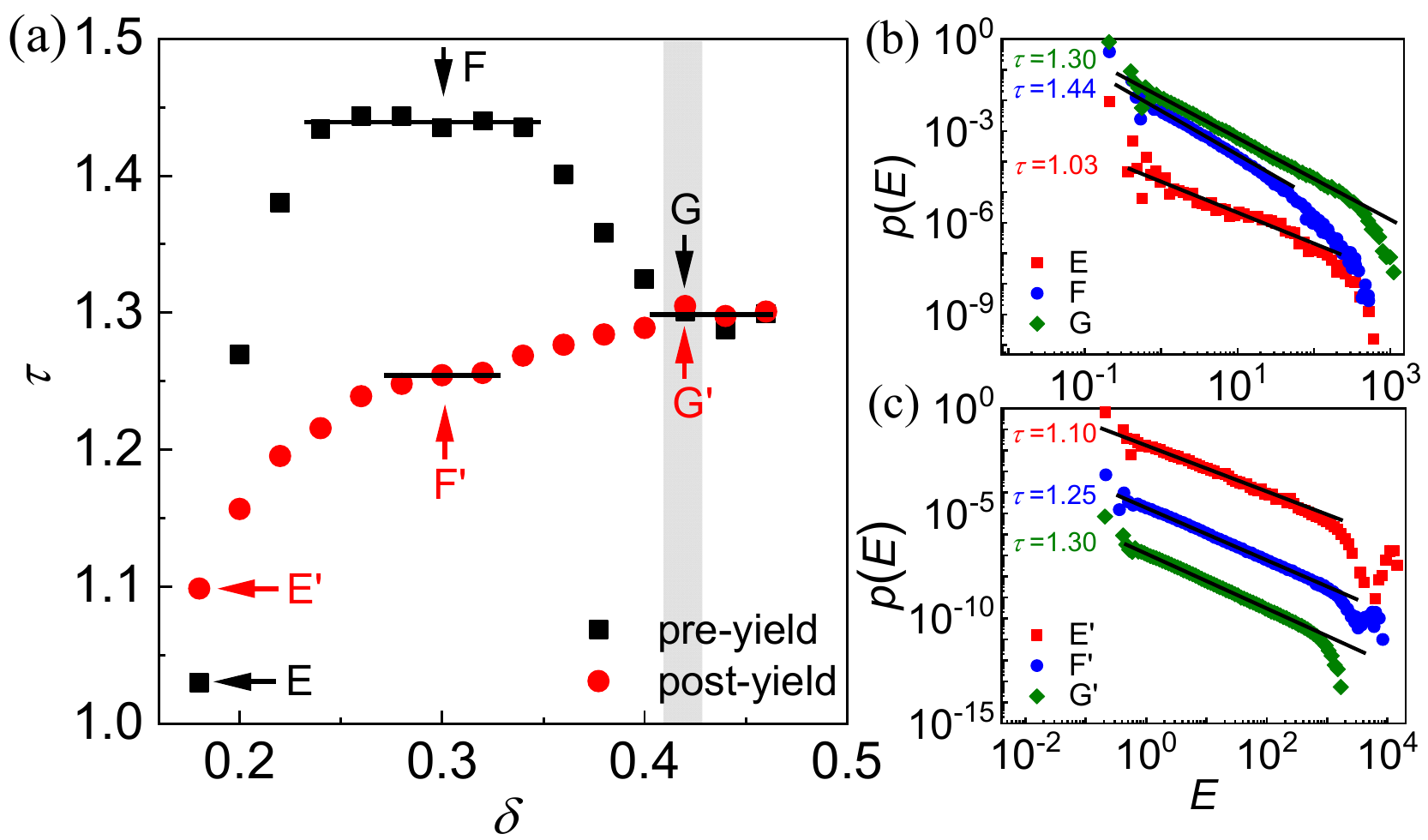}
\caption{  (a) Disorder dependence of the stress-resolved scaling exponent $\tau$ for immediately  pre-yield   and   post-yield regimes; (b,c)  corresponding  avalanche distributions for more than 100 realizations. The gray strip schematically marks the BD transition.\label{fig3}   }
\end{figure}

In  Fig.~\ref{fig3}(a) we show  the obtained continuous functions  $\tau(\delta) $  representing  pre- and post-yield exponents;  the associated  strain-resolved distributions are illustrated in Fig.~\ref{fig3}(b,c).

When  the  disorder is weak  ($ \delta \sim 0.2$),  the   pre- and post-yield exponents take almost the same   value  $\tau \sim 1$ (points $E$ and $E'$ in Fig.~\ref{fig3}(a)).  In such regimes,  where $R<1$,  homogeneously nucleated dislocations  are free to self-organize under the influence of  long-range elastic  forces \cite{Lehtinen2016-xc, Mordehai2018-qm} and the formation of a  shear band   only mildly affects  global dislocation dynamics.  The value of the exponent  $\tau \sim 1$  presents  a signature of  archetypically  'wild' plasticity  in the sense of  \cite{Weiss2015-eh}. 

The exponent $\tau =1$ has previously emerged in a fully analytical mean field theory of spin glasses  where  it  was associated with marginal stability \cite{ Pazmandi1999-lb, Franz2017-fa}. Based on this analogy one can argue that  around  $ \delta \sim 0.2$  our system generates sufficient self-induced  disorder to undergo a transition from  stable (elastic) to marginally stable  (or 'glassy' ) state whose phase space has a hierarchical (ultrametric) organization \cite{berthier2019gardner}. 
Such transition usually produces an almost   gap-less excitation spectrum \cite{Muller2015-bl} which we indeed see emerging in our system,  see Section \ref{SD}. The analogy with spin-glasses can be linked to the fact that during  yielding transition, the system effectively deals with only two neighboring energy wells of the infinitely periodic local energy landscape \cite{perez2008driving}. 

 The exponent $\tau =1$ was  also obtained numerically in the studies of quasi-elastic regimes in  structural glasses \cite{tyukodi2019avalanches,Ferrero2019-rx, Shang2020-zv}.   It was also found  to characterize dense amorphous packings and,   therefore,  can be associated with the concept of jamming.  In particular,
the avalanche exponent   $\tau =1$ is  predicted by  the fully analytical mean field theory for jammed packings \cite{Franz2017-fa}.

As we have already mentioned, the fact that the post-yield avalanche distribution in these regimes is super-critical,  see,  for instance,  our  data for the  500 nm Mo crystals shown  in Fig. \ref{fig1001}(b) and Fig. \ref{fig202}(b), is often not explicitly indicated  \cite{zaiser2008strain,cui2020role,Sparks2019-ie} even though it is well known that almost pure  nano- and micro-crystals always deform with  a system size  dislocational avalanche \cite{wang2012pristine,chrobak2011deconfinement,lu2011surface,
bei2008effects}.  The  super-criticality is also suppressed by   the neglect of dislocation nucleation in DDD simulations, even though  the exponent $\tau \sim 1$  emerges    in such models when they  rely on the assumption of  single slip plasticity and neglect disorder, e.g.  \cite{Ispanovity2014-ra}.   In fact, the authors of these studies have already   linked such regimes with  both dislocation jamming  and  self-induced  glassiness \cite{Ovaska2015-yb,Zhang2016-gh,Lehtinen2016-xc,ruscher2019residual}. 

%

\begin{figure}[!htbp]
\includegraphics[scale=0.8 ]{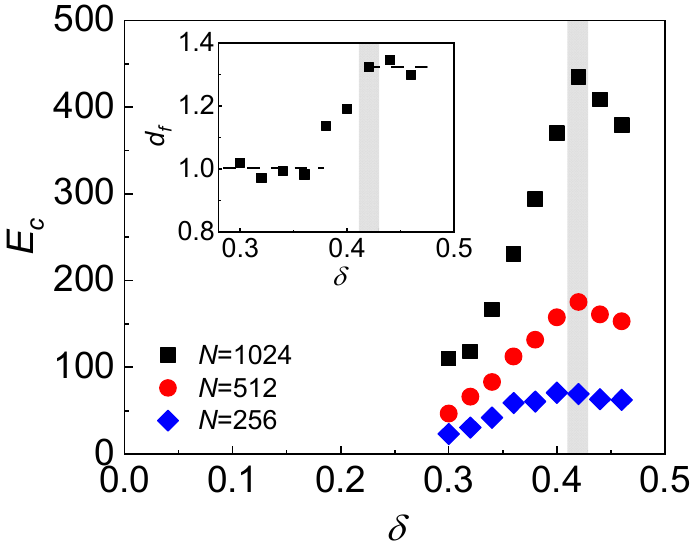}
\caption{ \label{fig4}   The divergence of the cut-off size in the  pre-yield regime around the critical value of  disorder $\delta \sim 0.42$. Inset shows the behavior of the fractal dimension $d_f$; the gray strip schematically marks the BD transition.}
\end{figure}

At the intermediate  disorder range, where $ 0.25<\delta < 0.35$,  
we observe a gap opening between the values of pre- and post-yield exponents (regimes $F$ and $F'$ in Fig.~\ref{fig3}(a)). In view of the progressive  rounding of the stress-strain curve near the  \emph{strain}
 controlled spinodal point,  where  \cite{Popovic2018-pp}
$
d\sigma/d\gamma=-\infty,
$
  one can expect   the pre-yield scaling  to represent the  spinodal nucleation which shows scale-free features due to  the dominance of long-range elastic interactions  \cite{ Ferguson1999-px, Nandi2016-lb,Procaccia2017-gq,da2020rigidity}. In the post-yield regime, the pdf's characteristic peak still indicates  nucleation of system size shear bands, while the scale-free range can be linked to their \emph{spreading} in the form of elastic depinning 
 \cite{Ovaska2015-yb}.  A prototypical  example of a double-well system with a long-range spinodal,  where nucleation and propagation (depinning)  exponents  are different,  is discussed in \cite{perez2008driving}.
  
 Around  $ \delta \sim 0.42$   the first order phase transition terminates in a critical point representing the  BD transition and the  pre- and post-yield exponents collapse again (regimes $G$ and $G'$ in Fig.~\ref{fig3}(a)).  Between  $ \delta \sim 0.3$ ( point $F$   in Fig.~\ref{fig3}(a)) and  $ \delta \sim 0.4$ (regime  $G$   in Fig.~\ref{fig3}(a))  the pre-yield exponent $\tau$ exhibits a characteristic disorder-induced crossover from spinodal to critical scaling discussed for  the mean field setting in \cite{da2020rigidity}. 

In the critical BD regime at $ \delta \sim 0.42$  the  scaling exponent takes the value  $\tau \sim 1.3$,  which  is close to the one observed  for slip-size statistics   in  nano-pillars  (both FCC and BCC)    \cite{Dimiduk2006-fz,Brinckmann2008-od}; the same value   characterizes plastic yield in  amorphous solids \cite{salerno2013effect,lin2014scaling,Liu2016-tw,Budrikis2017-ex}. In a DDD model with quenched disorder similar value of the exponent $\tau$ was obtained in \cite{Ovaska2015-yb}. 
 
The fact that around   $ \delta \sim 0.42$ we encounter  a critical point   was already  hinted upon by  our finding of the  multi-fractal structure of the plastic strain field at this strength of disorder, see Fig. \ref{fig41}. To provide additional evidence,  we show  in Fig.~\ref{fig4} the disorder dependence of the cut-off  parameter $E_c$ characterizing    stress-integrated  pre-yield avalanches.  It peaks at the critical value of disorder $ \delta \sim 0.42$, where  it diverges with  system size,   $E_c\sim N^{d_f}$. The  fractal dimension of avalanches $d_f$ jumps during the BD transition  from the value $\sim 1$, in the brittle phase \cite{Csikor2007-jk, Ispanovity2014-ra},  to the  value $\sim 1.4$  in the ductile phase. Note that  the latter  is close to the computed fractal dimension of the strain pattern at this level of disorder, see Section \ref{SC}.

The BD criticality  in this problem emerges within a broad  range  of $\delta$, which is not uncommon for  systems  with long-range  correlations,  where  the presence of rare but strong spatial disorder fluctuations can divide  the system into spatial regions which independently undergo the  transition \cite{vojta2006rare}.  For   a   \emph{subsystem},  characterized by the (average) stress-strain relation $\sigma(\gamma)$ and  effectively loaded through an  elastic matrix   with   stiffness $\mu$, the  condition of criticality  would be
$
 d \sigma/d \gamma=-\mu,\,\  d^2\sigma/d \gamma^2=0.
$
The  power-law distributed avalanches  can   be   then generated in  sub-systems  exposed to  different  $ \mu \in (0,\infty) $, even though the 
 macroscopic critical point  formally corresponds only to $\mu=\infty$.

As the strength of the disorder increases beyond  $ \delta \sim  0.5$,  plastic hardening takes over starting  at  almost zero stress, and scaling is  getting lost.  Instead of large-scale heterogeneous avalanches,  the model shows   homogeneous proliferation of uncorrelated plastic activity,  much  like what we see in  the experiments  on   1500  nm samples,  see Fig.~\ref{fig1001}(b) .


\begin{figure}[!htbp]
 	\includegraphics[scale=0.6]{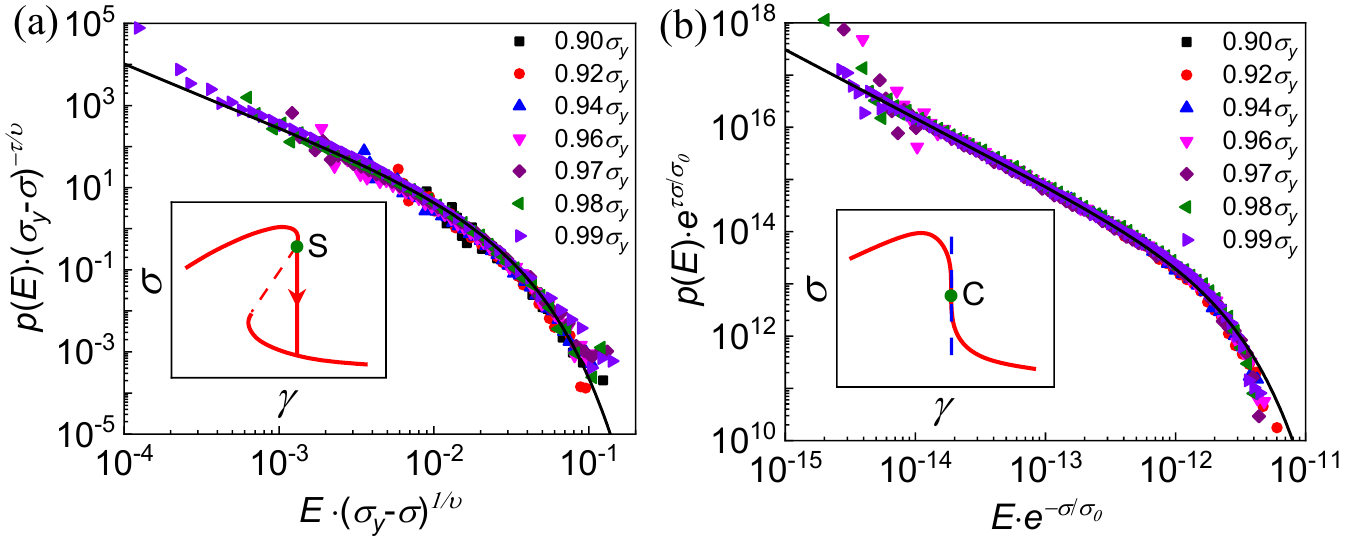}
 	\caption{ Scaling collapse for the two cases:  (a) $\delta =0.30$ (tuned spinodal criticality)    and (b) $\delta =0.46$   (BD criticality). Insets show schematic   stress-strain curves around spinodal $S$ and critical $C$ points. \label{fig10}  }
 \end{figure}

 Finally we show the difference in the nature of the scaling collapse  of pre-yield data shows for  the critical and the spinodal points.
In all near-critical regimes, the   stress-resolved energy distribution  is  of the form 
$
p(E;\sigma) \sim E^{-\tau}f(-E/E_c(\sigma)) 
$
and to obtain the functions $f$ and $E_c$, we  need to re-plot our data  using  the normalized variables $E^{\prime}= E/E_c$ and $p^{\prime}= p(E) E_c^\tau$.  We find  two distinct regimes where such  data collapse could be achieved.

\begin{figure}[!htbp]
 	\includegraphics[scale=0.6]{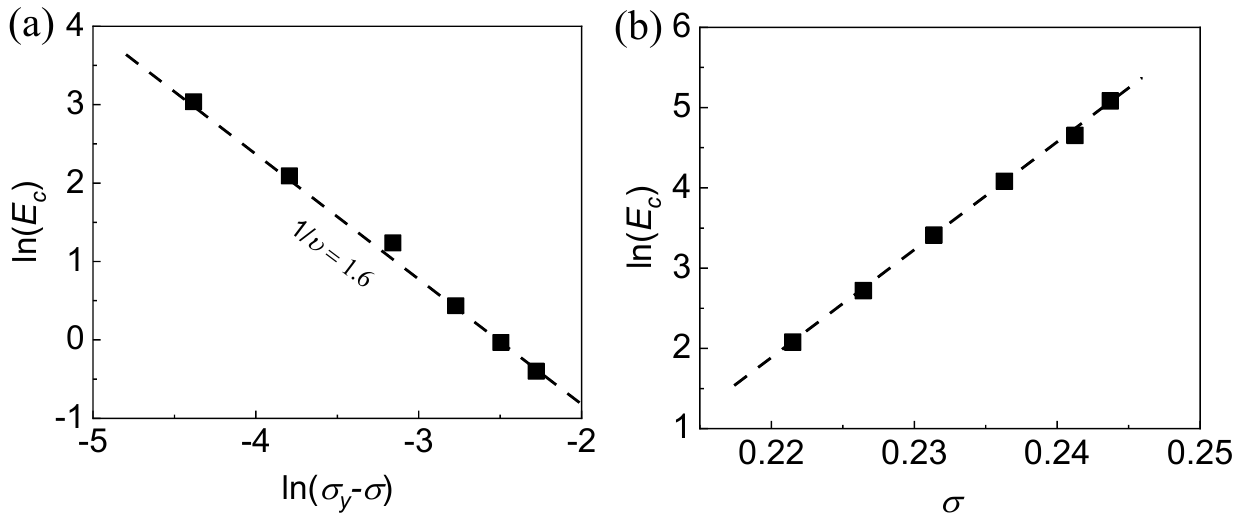}
 	\caption{Critical scaling for:  (a) $\delta =0.30$ (tuned spinodal criticality) and (b) $\delta =0.46$   (extended  BD criticality).  \label{fig110}. }
 \end{figure}

In the   interval $0.3 < \delta < 0.4 $,   our   Fig.~\ref{fig110}(a)  shows  the validity of  the  scaling ansatz in the cutoff region with  
$
E_c (\sigma) \sim (\sigma_y - \sigma)^{-1/v}.
$
Here $\sigma_y$ is the yield stress at $\gamma_y$, and $v$  is a constant. This scaling suggests tuned criticality; indeed, the spinodal point is associated not only with a particular strain but also with particular stress \cite{procaccia2017mechanical,da2020rigidity}.   We  show  in Fig. \ref{fig110}(a)  that $1/v \approx 1.6$, which is different, for instance,  from the value predicted in the theory of mean-field depinning where  $1/v=2.0$ \cite{dahmen2009l}, see Section \ref{LN} for the relevance of this comment.  

\begin{figure*}[!htbp]
\includegraphics[scale=1.1]{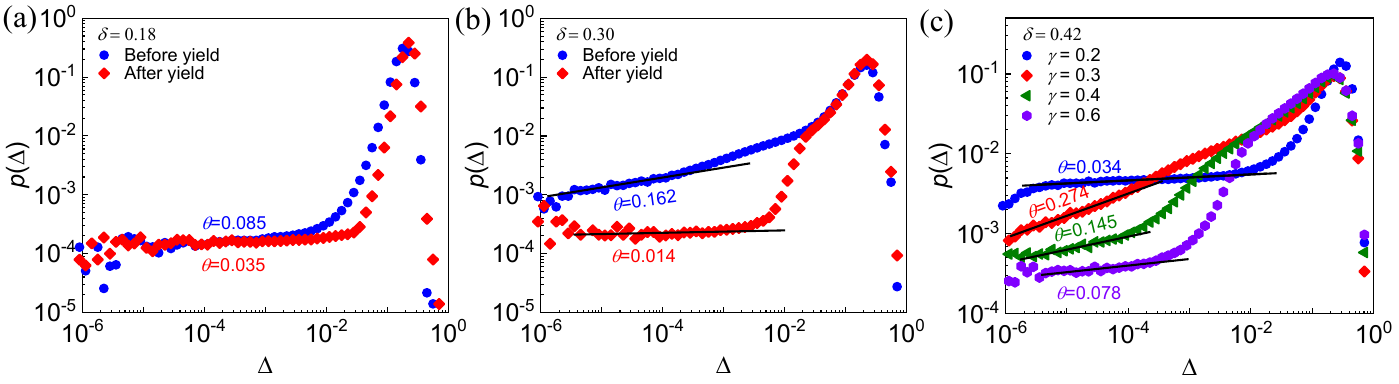}
\caption{Distribution  of   stability  measures  $p(\Delta)$ at: (a)  $\delta=0.18$ (weak disorder), (b) $\delta=0.30$ (intermediate disorder) and (c) $\delta=0.42$ (critical disorder).  In (c) we effectively show the function of two variables $p(\Delta; \gamma)$. 
\label{fig88}   }
\end{figure*} 
The second region of scaling collapse is  around  $\delta \sim 0.42-0.46$.  Here  the cutoff follows  the asymptotics
$
E_c (\sigma) \sim \exp(\sigma/\sigma_0),
$
where $\sigma_0$ is a constant. The absence of stress tuning in this case can be explained by the fact  that criticality in  a strain-control ensemble  makes   stress poorly constrained,  see  Fig.~\ref{fig110}(b). 
In other words,  the BD critical point  at $\delta \sim 0.42$ is localized in strain but not in stress.   
 

\section{ Excitation spectra}
\label{SD}

Recent advances in amorphous \cite{lin2014scaling} and crystal plasticity \cite{ovaska2017excitation}  suggest that an important 
 characterization of  threshold-controlled  dynamics  comes from  the  density of elements with a given level of stability,  also known as the excitation spectrum.
 
 In our problem the  natural   stability measure  is  $\Delta =  \bar\xi^e-\xi^e$ where $\xi^e=\xi-d(\xi)$ is the elastic  strain and  $\bar\xi^e=0.5$ is the  stability threshold. The excitation spectrum is then  the  distribution $p(\Delta)$  of the local distances   to a
threshold above which a plastic correction takes place.

The form of the excitation spectrum in the limit $\Delta \to 0$ can be linked to the nature of intermittent fluctuations  exhibited
by the system \cite{Lin2015-go, lin2016mean}.  Of particular interest is  the “pseudogap” exponent $\theta(\gamma,\delta)$  entering the asymptotics 
$
p(\Delta) \sim \Delta^{\theta}
$ at $\Delta \to 0$ \cite{karmakar2010statistical,Muller2015-bl}.

We recall that  in the case of    classical  depinning of an elastic manifold  in random media, when  yielding of a given site can only increase the load of other sites, it was shown that $\theta=0$  \cite{fisher1998collective}.  Another paradigmatic case is plasticity of amorphous glasses  \cite{lin2014scaling,shang2020elastic}  where $\theta>0$ which was linked to the fact that the elastic long range interaction kernel is sign-indefinite. 

While  the depinning remains one of the main paradigms of plastic yielding in crystals \cite{Friedman2012-ie, Ovaska2015-yb},  it was also realized  stress transfer in crystal plasticity is  sign-indefinite  \cite{bako2007dislocation,Ispanovity2014-ra, Lehtinen2016-xc}, and  an extensive study  \cite{ovaska2017excitation}    produced a  singular  excitation spectrum with the  exponent $\theta>0$ showing  dependence on quenched disorder. The excitation spectra generated in our  numerical experiments are summarized in Fig. \ref{fig88}.  

When the quenched disorder is weak, see  Fig. \ref{fig88}(a), the pre- and post-yield  exponents $\theta$ agree. The  obtained   spectrum is almost  gap-less  with very small value of $\theta>0$. This is an indication of   weak criticality  \cite{Hentschel2015-pj,tyukodi2019avalanches,ferrero2019criticality,Franz2017-fa,Shang2020-zv} when   the probability to find infinitesimal energy barrier is  finite but close to zero. In such states  the system is close to being  elastic  with dislocation microstructures characterized by high energy and low  stability. Such  systems are usually  'marginally stable'  in the sense that  instabilities  start to  occur as soon as infinitesimal  extra loading is applied, with  glasses  near and above jamming point as a prominent example \cite{Muller2015-bl}. Based only on the value of the exponent $\theta\sim 0$ it would be  difficult  to distinguish jamming from depinning in this case,  however, since  we know that $\tau \sim 1$ the jamming scenario should be  clearly favored \cite{Ispanovity2014-ra}. 

The excitation spectrum with  $\theta\sim 0$ was also recorded in some mesoscopic models of amorphous plasticity \cite{ferrero2019criticality,tyukodi2019avalanches} and molecular dynamic simulations of glasses \cite{shang2020elastic}. As we have already mentioned in Section \ref{AS},  this overall behavior  is  similar to the marginal response of mean-field systems described by the replica symmetry breaking  framework and   is also  in  agreement  with what was found in simulations of three-dimensional systems of soft spheres, either at jamming  or at slightly higher densities \cite{Franz2017-fa}.

At the intermediate level  of disorder $\delta \sim 0.3$, see  Fig. \ref{fig88}(b), we observe  the emergence of a stronger pseudo-gap in pre-yield conditions,  which we interpret as a signature of spinodal criticality.  Instead, the post-yield regimes in this range of $\delta$ are characterized  by   $\theta \sim 0 $.  This  may be explained by elastic depinning of an advancing  surface  separating  a  shear band   from  the rest of the crystal. Such surface   would be   generically  produced by a   spinodal  SNAP event and its intermittent dynamics will be then  controlled by  sign-definite  surface elasticity \cite{perez2008driving}.

 Around the  BD critical point at $\delta \sim 0.42$ we observe a non-monotone dependence of the exponent $\theta$  on the loading parameter with a strong maximum around the yield strain $ \gamma_y$, see  Fig. \ref{fig88}(c).   Before the yield,    we obtain $\theta \sim 0$  which can be  again  interpreted as a signature of dislocation  jamming.  The yield at this level of disorder comes with  an opening of a pseudo-gap  indicating  the  development  of  the global connectivity of the energy landscape  \cite{Muller2015-bl,zhang2017scaling}. The increase of the applied strain  $\gamma$  beyond the critical strain,  decreases the pseudo-gap exponent again,  bringing it to  a plateau with   $\theta \sim 0$,   characterizing the  stationary
regime \cite{ruscher2019residual}. We can conjecture that this as a signature of the emerging depinning scaling. Indeed,   in such  regimes the  self-induced   inhomogenity,  due to dislocation entanglements,   can be expected to compromise  the  sign-indefinite nature of the elastic kernel with  long range interactions  progressively getting replaced  by the largely ferromagnetic,  short range interactions.  
 
At even stronger   quenched disorder  ($\delta > 0.6$ )   one can expect the  loss of scaling and proliferation of  uncorrelated POP events.

\section{Mean field model}
\label{MF}

 A simple  mean field  model  can be used to rationalize  at least some elements of the observed behavior in terms of  macroscopic parameters.  
Suppose  that the stress resolved evolution of the spatially averaged density of  mobile dislocations  $\rho$  is described  by a stochastic kinetic equation   \cite{Weiss2015-eh}
\begin{equation}
 \rho^{-1}d\rho/ d\gamma= -c+\sqrt{2D}\eta(\gamma),                                         
 \label{stoch}
\end{equation}
where  the local shear strain $\gamma$ serves as a time-like parameter,  $c \geq 0$ characterizes the  rate of dislocation immobilization and   the  
temperature-like parameter $D$ represents the intensity of the multiplicative mechanical noise with  $\langle \eta(\gamma) \rangle=0$ and $ \langle \eta(\gamma_1 ),\eta(\gamma_2) \rangle=\delta (\gamma_1  -\gamma_2)$. Note that the lack of  conventional  dislocation sources in sub-micron crystals allows us to  neglect  here the  Kocks-Mecking  dislocation nucleation  term \cite{kocks2003physics,Weiss2015-eh}. 

\begin{figure}[!htbp]	
\includegraphics[height=6cm]{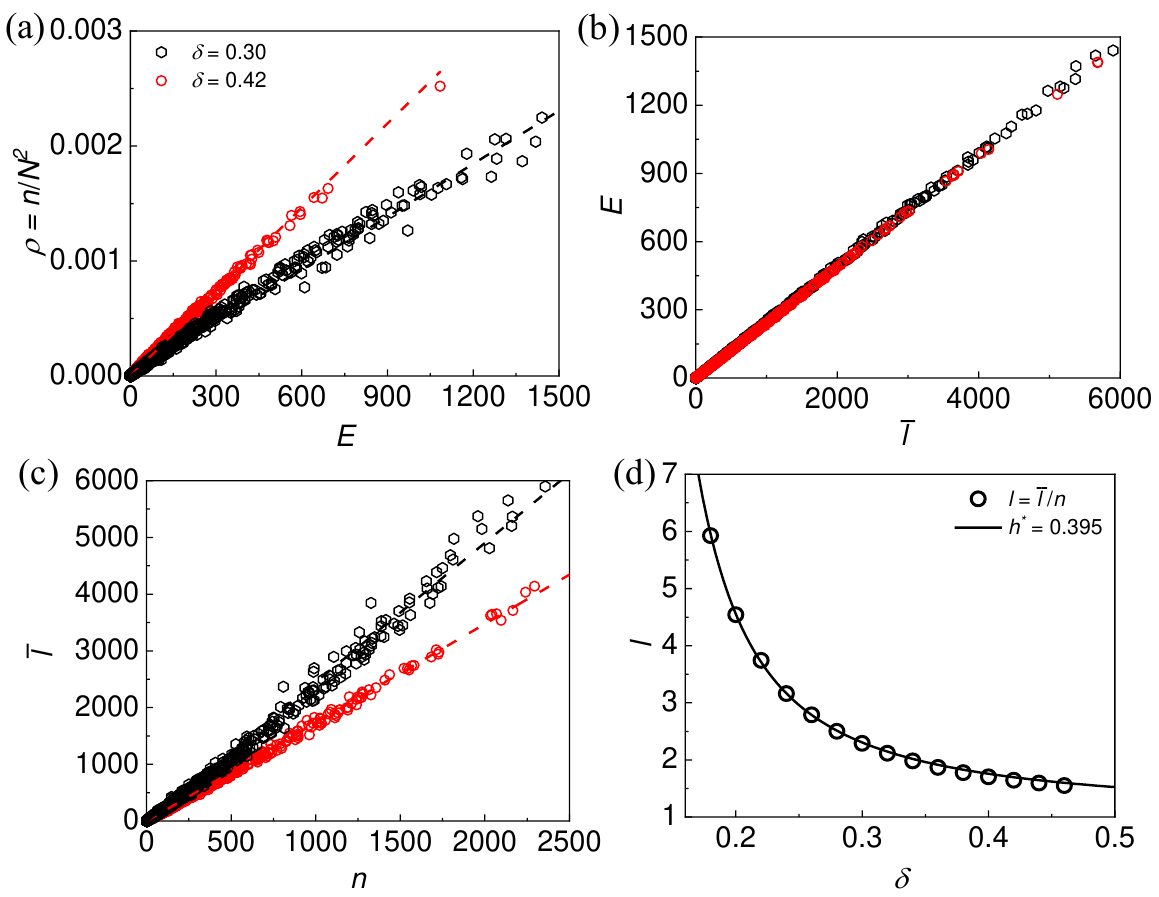}
\caption{ (a) Relation between the density of mobile dislocations   $\rho$ and avalanche energy $E$ for representative disorders $\delta=0.30, 0.42$; (b) dependence of  $E$  on  the  total  distance covered by  dislocations   during an avalanche $\bar l$; (c)   dependence of $\bar l$ on the number of  moving   dislocations   $n$; (d) Internal length scale $l$  as a function of disorder strength $\delta$. \label{rho_E}}
\end{figure}

 The   stationary probability distribution in  \eqref{stoch} is of a pure power law  form 
$
 p_s (\rho)   \sim  \rho^{-\alpha}
$
with the exponent $\alpha=1+c/D$.  In the framework of our automaton  we can interpret  $\rho$ as the density of mobile dislocation during an avalanche at a given value of the loading $\gamma$.  We can then write $\rho(\gamma)=n(\gamma)/N^2$,   where $n(\gamma)$ is the number of dislocations moved during  an  avalanche.  Our numerical experiments suggest that the avalanche energy $E$ is a disorder independent linear function of the total distance traveled by mobile dislocations during an avalanche $\bar  l$, see  Fig.~\ref{rho_E}(b), and that  $\bar l \sim n$, see    Fig.~\ref{rho_E}(c). Therefore  $E \sim \rho$, see  Fig.~\ref{rho_E}(a), and   we can conclude  that  the exponent $\alpha$ in the mean field model  is the same as  the exponent $\tau$ in the automaton model.  The relation $\alpha=\tau$, relying on the fact  that for nano-crystals   the mean free path is controlled only by the strength of quenched disorder (a proxy of the crystal size),   is not applicable to bulk materials where one can expect that  $\alpha=2\tau-1$ \cite{Weiss2015-eh}. 

For  single slip  pure  nano-crystals with weak disorder dislocation immobilization  can be neglected,   so  $c/D \ll 1$, and   the stochastic evolution of  $\rho$ governed by  \eqref{stoch} reduces  in this case to  a geometric Brownian motion with $\alpha\sim  1$. In the automaton model we  observe in the  low-disorder  limit dislocation self-organization,  governed exclusively by elastic long-range elastic interactions \cite{Ispanovity2014-ra, weiss2019ice},  and recover the same value of the exponent $ \tau \sim  1$.   With increasing disorder, the immobilization rate $c$ should increase leading to a higher value of $\tau$,  which is in qualitative agreement with our numerical experiments.
 
 The  crossover from  $D$-dominated brittle regimes  ($c<D$ with stochastic term in  \eqref{stoch} controlling dynamics) to  $c$-dominated ductile regimes ($c>D$ with deterministic term in  \eqref{stoch} controlling dynamics) can be expected where the  mechanical agitation is balanced by dislocation self-locking  ($c \sim  D$).  Even though our oversimplified model \eqref{stoch} is not designed to capture the avalanche distribution in ductile regime, it is of interest to use our numerical results for reformulating the above limits in terms of crystal sizes.
\begin{figure}[!htbp]
\includegraphics[scale=0.2]{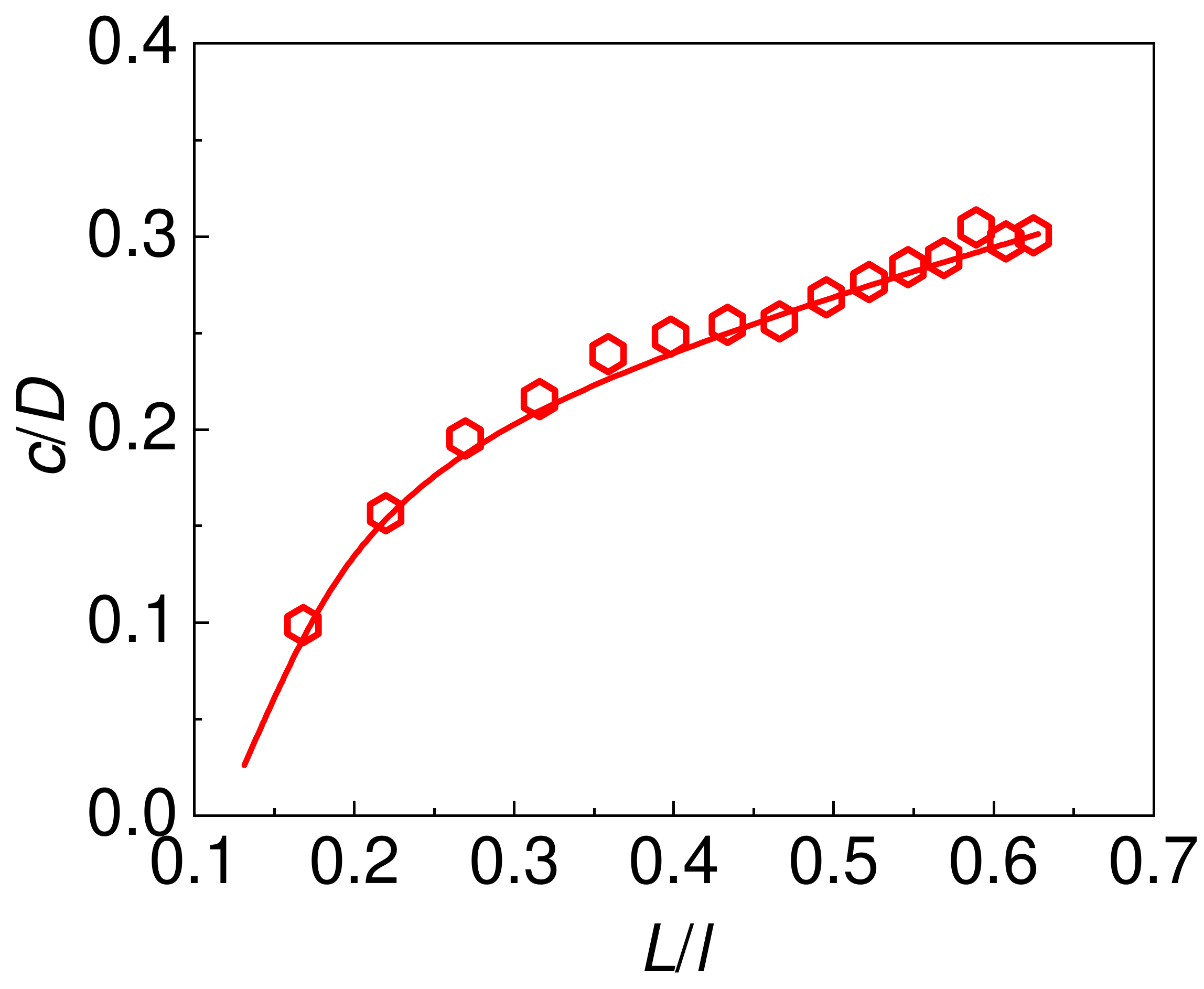}
\caption{  Dependence of the exponent $c/D$ in the post-yield regime on the dimensionless parameter $R=L/l$  for   $h^*=0.395$ and $L=1$
\label{fig81}   }
\end{figure}

 Consider  first  the mean free path $l=\bar l/n$, characterizing  dislocation glide before it gets immobilized.  In  Fig.~\ref{rho_E}(d) we show its dependence of disorder strength $\delta$ obtained from our numerical experiments. To obtain an analytical relation we can assume   that only defects with the  strength above some threshold $h^*$ participate in immobilization and that they form a regular lattice with spacing $l$. Then we can write  \begin{equation}
L/l \sim ( 1- \operatorname{erf} (h^*/(\sqrt{2}\delta)))^{1/2}.
\end{equation}
Our  Fig.~\ref{rho_E}(d) shows that this relation provides a perfect fit for  the empirical curve  if  the stress threshold  takes the value $h^*=0.395$. It also suggests  that in the automaton model  the mean free path of dislocations $l$,  setting an intrinsic internal length scale, is  indeed  controlled by the tails of  the disorder distribution.
 
Given that we know the relation $\tau(\delta)$ for  post- yield regimes from our numerical  experiments (see Fig. \ref{fig3}(a)), and using the relation $\alpha=\tau$,  we can now obtain a relation   between $c/D$ and  $R=L/l$, see Fig. \ref{fig81}. It provides  the desired \emph{quantitative} description of  the    crossover from  brittle (nano-crystalline) to  (ductile) micro-crystalline   plasticity. 

Indeed,  the effective temperature $D$ should depend only weakly  on  the system size $L$.  It is defined instead  by the  locking strength of defects, which means that it increases  with $ l$. At the same time, it is clear that  the rate of dislocation reactions (in particular our parameter $c$  controlling immobilization)   increases   with $L $ \cite{Zhang2017-cl}. Therefore,    in either very small  and/or  very weakly disordered samples   $c<D$ and the response must be brittle.  Conversely,  in either bigger or more disordered  samples one  can  expect to reach the  ductile phase  where $c>D$.  This is  what  the curve shown in Fig. \ref{fig81} implies.


\section{Cyclic loading}
\label{CL}

The mechanical response to  monotone loading   carries a memory of the initial state and  in our tests, the  preparation was  entirely  dislocation-free, as the goal was  to simulate the plastic deformation of ultra-small systems (nano-particles and nano-pillars) \cite{Sharma2018-iw,Mordehai2018-qm}. To obtain the  generic  response, one can  prime  the crystal  by subjecting it  to   cyclic protocol.  If the  strain amplitude  of such pre-loading extends  beyond the yield,  a crystal    becomes dislocations-rich  already  after a first cycle,  independently of the  initial  disorder.

 \begin{figure}[!htbp]
	\centering	
	\includegraphics[scale=.65]{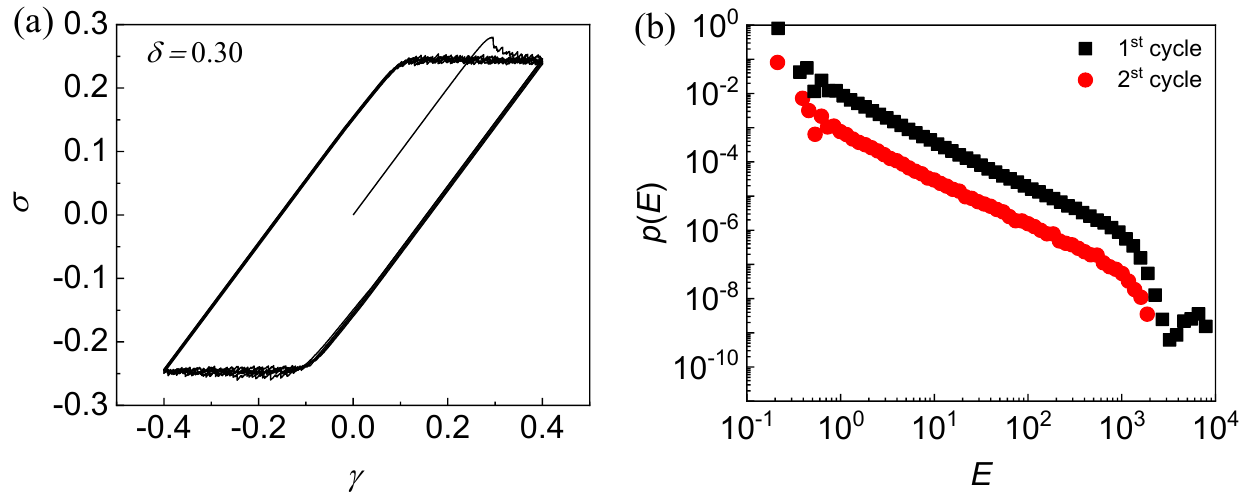}	
	\caption{  (a) Strain-stress curves for the crystals showing six loading-unloading cycles; (b)  Avalanche distributions of cyclically loaded crystals during the first and the second cycles; the first cycle here is understood as a  monotone loading path. Here:  $\delta=0.30.$
		\label{cyclic_distribution1}}
\end{figure}
 
The resulting  self-induced disorder can be expected to quickly overtake the quenched disorder. As a result, the dimensionless parameter $R$ will  increase  due to the  decrease of the dislocation mean free path $l$, which is now controlled by the density of the generated dislocations. This will lead to mild ductility ($R>1$) of even sub-micron crystals  but without  strain-hardening because of the remaining  single-slip arrangement of   plastic  flow.

The typical stress-strain curves generated by the   automaton model in  quasi-static cyclic  loading conditions are illustrated in Fig.~\ref{cyclic_distribution1}(a). 
Brittleness indeed disappears  already after the first load reversal   even for the case of relatively weak  initial  disorder.  The  decrease in yield stress is consistent with the observations of    softening in nano-crystals in response to  the  increase in the  number of initial dislocations \cite{bei2008effects}.  

\begin{figure}[!htbp]
	\centering	
	\includegraphics[scale=.25]{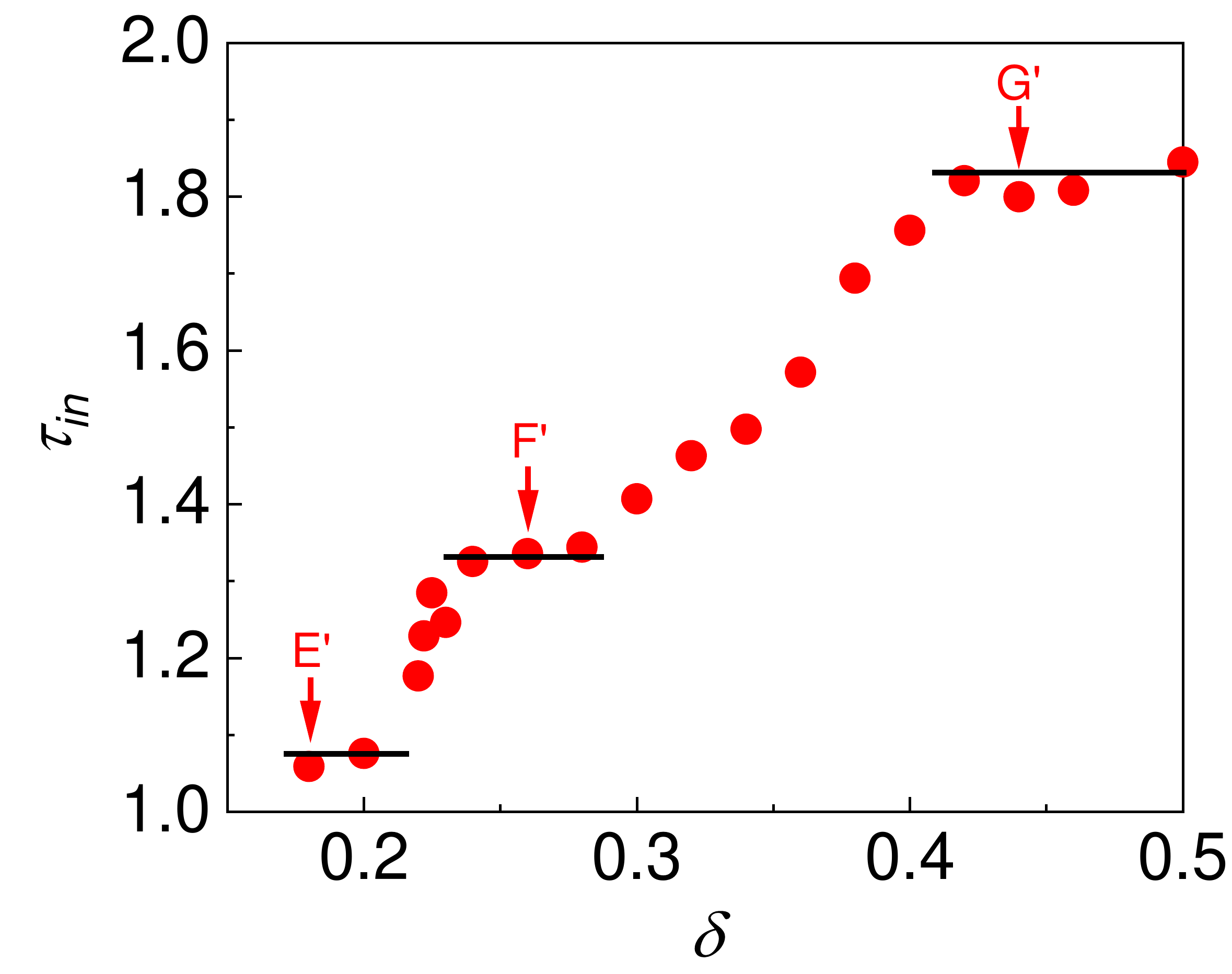}	
	\caption{ Disorder dependence of the integrated exponent $\tau_{int}$ in the case of cyclic loading. 
		\label{cyclic_distribution}}
\end{figure}

In Fig.~\ref{cyclic_distribution1}(b) we show how a typical  avalanche distribution  changes after the first cycle for  $\delta= 0.30$. One can see that the originally super-critical avalanche distribution, involving characteristic system size events, is replaced by a near-critical distribution that  is basically maintained with subsequent cyclic loading.  A robust  range of scale-free behavior with a stable  exponent $\tau$ emerges after  shakedown. This observation suggests that a strongly ductile regime,  with $R \gg 1$ and the manifest loss of scaling, cannot be achieved by cyclic loading at least at this strength of quenched disorder.

In Fig. \ref{cyclic_distribution} we show the disorder dependence of the stabilized cycle-integrated exponent $\tau_{in}$ which emerges in the case of sufficiently large amplitude of cyclic loading.  The $\tau_{in}(\delta)$  curve shows  distinctly three characteristic plateaux corresponding to the same three main scaling regimes which we  identified in Section \ref{AS} and can conditionally label as glassy, spinodal and critical.  While the  values of the exponents in Fig. \ref{cyclic_distribution} and in Fig. \ref{fig3}(a) do not match exactly due to the different nature  of these exponents   (stress-resolved vs. cycle-integrated) and   different conditions (cyclic vs. monotone loading),  the main trends  appear to be well maintained.

Thus, at low strength  of quenched disorder (plateau around point $E'$) we obtain the value of the exponent  reminiscent of the one in mean field spin-glasses  and suggesting that  dislocations can  self-organize into a marginally stable (jammed) state. At the intermediate level of disorder (plateau around point $F'$)  we see the scaling  which we previously associated with spinodal  criticality when a SNAP even produces a system size effective manifold which evolves through  classical elastic depinning. Finally, at sufficiently  large strength  of quenched disorder (plateau around point $G'$), we observe a  stretched scaling range associated with the critical BD transition.  

In this analysis, we have identified  two major crossover regimes. The first one,  from $E'$ to  $F'$,  is similar to the jamming-to-depinning transition  studied numerically by DDD approach in \cite{Ovaska2015-yb}. The second   one, from   $F'$ to  $G'$, is    similar to the spinodal-to-critical transition  studied analytically at the mean field level in \cite{Ozawa2018-xi,Popovic2018-pp, da2020rigidity}.  While in previous work  these crossovers were associated exclusively with changing disorder, here we  interpret them as a feature of a size effect.

Finally, we note that our observation about the disappearance of brittleness in cyclic loading can potentially be used to turn fragile nano-crystals into mildly ductile nano-crystals \cite{papanikolaou2017avalanches}.  Given the vulnerability of brittle ultra-small structures,  the possibility to enhance ductility  by purely mechanical means  is  of considerable interest for applications \cite{Zhang2017-cl}.  Moreover,  by effectively increasing the strength of quenched disorder, such 'training'   can be expected to bring the crystal  closer to the critical state \cite{Perez-Reche2016-xp}.

\section{Compatible  disorder}
\label{LN}

In addition to disorder, characterized by the random function $h_{i,j},$ and representing pre-stress acting directly on the longitudinal strain variable $\zeta_{i,j}$ and indirectly on the shear strain variable $\xi_{i,j}$, one can also  introduce a  disorder-related pre-stress $g_{i,j}$ acting directly on  $\xi_{i,j}$ \cite{Tang2020-aj}. Then the energy density takes more symmetric form \cite{ Salman2011-ij,Salman2012-oa}:   
 \begin{equation}
 f (\xi_{i,j},\zeta_{i,j} ) = \frac{K}{2}\zeta_{i,j}^2+\frac{1}{2}(\xi_{i,j}-d_{i,j}(\xi))^2-h_{i,j}\zeta_{i,j}-g_{i,j}\xi_{i,j}.
 \end{equation}
 
 Note that in contrast to $h_{i,j},$ the disorder $g_{i,j}$  acts on  the primary  order parameter variable   locally  as in the conventional  RFIM \cite{Nandi2016-lb}. Such 'local' disorder can be viewed as resulting from lattice-compatible obstacles inhibiting or promoting plastic slip only in the narrow vicinity of a compact source of the disorder. One can think, for instance,  about   locked dislocation multi-poles, whose long-range fields are effectively screened. 'Local' disorder may also be  related to lattice-scale  inhomogeneities lowering or raising the  Peierls stress point-wise. 
 
Suppose that  both disorder  fields, $h$ and $g$,  are   drawn independently in each lattice cell from  Gaussian distributions   
\begin{equation}
p_{s}(r)= (2\pi\delta_s^2)^{-1/2}\exp{(- r^2/(2\delta_s^2))},
 \end{equation} 
 where $s=(g,h)$.  In the previous Sections we used the special notation   $ \delta \equiv \delta_h $ but here, to distinguish the two,  we'll keep the notation $\delta_h$. 
 
The   specificity of the disorder $ g_{i,j} $, representing essentially  a residual plastic strain,   is that it can be  simply combined in the energy density with the actual plastic strain $ d_{i,j}$. For instance, to account for $g$ in the Fourier representation of the elastic solution,  it sufficient to  replace the  field $\hat  d(\bold q)$  by the   sum $\hat  g(\bold q)+\hat d\bold(\bold q)$.  We can then write 
\begin{equation} 
 \hat\xi(\bold q) = \gamma\delta(\bold q) + \hat L (\bold q) \left[\hat d(\bold q) +\hat g(\bold q)\right]  + \hat L_h (\bold q)  \hat h(\bold q), 
\label{automaton3}\end{equation}
where the kernels  $\hat L_h (\bold q)$ and  $\hat L_h (\bold q) $ were introduced in \eqref{kernel1} and \eqref{fk2}, respectively.
 
To perform a direct comparison of the  two types of disorder  we need to assess the  action of the fields $g_{i,j}$ and $h_{i,j}$  on the \emph{same} strain variable.  A natural way to do this is to  eliminate the linear non-order parameter  $\zeta_{i,j}$ adiabatically and to evaluate  the role of disorder $h_{i,j}$ in the 'condensed' model containing only one variable  $\xi_{i,j}$. The crucial observation is that the strain variables   $\xi_{i,j}$  and  $\zeta_{i,j}$  are not independent even though they are not coupled explicitly in the energy density.   The implicit coupling can be revealed if we recall   the  constraint of geometric compatibility ~\cite{shenoy2008strain}. 

\begin{figure}[!htbp]
	\centering	
	\includegraphics[scale=0.2]{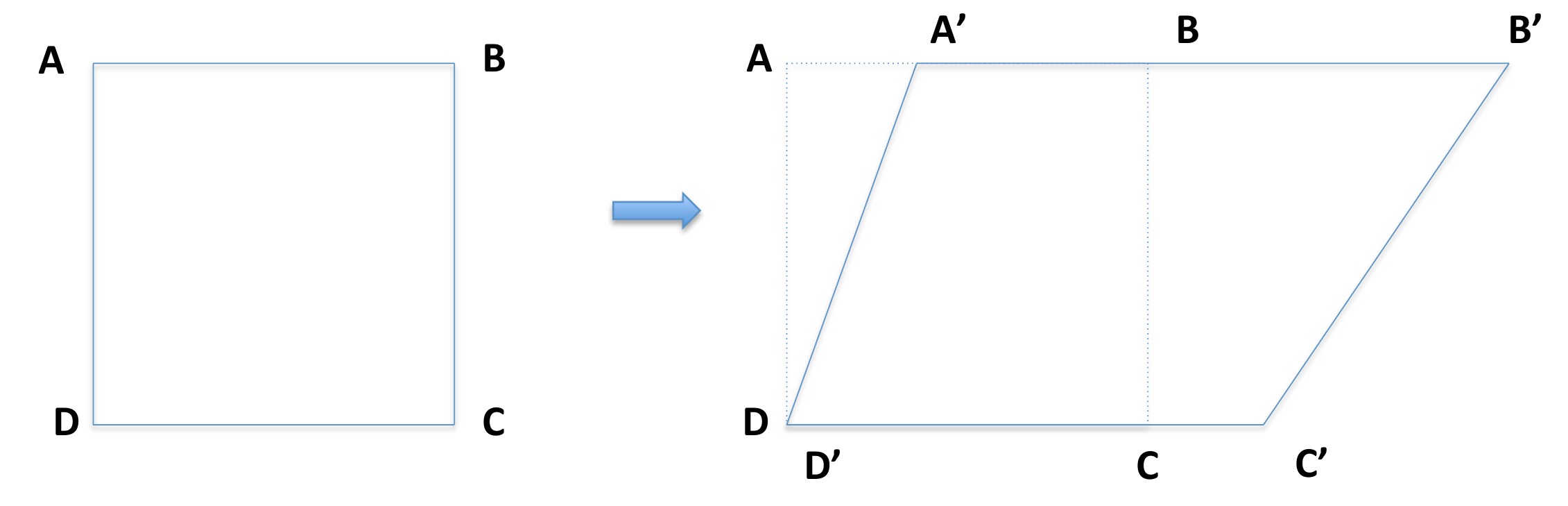}	
	\caption{Generic deformation of a mesoscopic square element.
		\label{peng}}
\end{figure}

From now on, it will be  convenient to deal directly with physical variables rather than their  Fourier transforms. Since our displacement field is scalar,  the  generic  deformation of an element is  highly anisotropic, see Fig. \ref{peng}.  The geometrical meaning of the  two strain variables $\xi_{i,j}$  and  $\zeta_{i,j}$ becomes   apparent from the identification: $D'C'=1+\zeta_{i,j}$, $A'B'=1+\zeta_{i,j+1}$, $AA'= \xi_{i,j}$, and $BB'-CC'= \xi_{i+1,j}$, where $ABCD$ is the square lattice element before the deformation and $A'B'C'D'$ is its image after the deformation, see Fig. \ref{peng}. 

It is straightforward to see that 
$
AA'+A'B'=1+BB',
$
and since $D'C' -CC'=1$ we obtain in terms of  $\xi_{i,j}$  and  $\zeta_{i,j}$ \cite{Perez-Reche2016-xp} 
\begin{equation}
\xi_{i,j}-\xi_{i+1,j}=\zeta_{i,j}-\zeta_{i,j+1}.
\label{stvenant}
\end{equation}
Eq. \ref{stvenant} is a  discrete analog of the classical Saint-Venant compatibility relations in continuum linear elasticity, see  \cite{treibergs2020compatibility} for the general analysis. It provides a   constraint on the variables $\xi_{i,j}$  and  $\zeta_{i,j}$ which is inherently nonlocal. If we   now complement   the  condition \eqref{stvenant}  with  our  (single)  mechanical equilibrium condition \eqref{equil}, we obtain a closed system representing in  our case the  classical Beltrami-Michell equations of classical elasticity \cite{kucher2004some}.

To simplify notations we'll be using  lexicographic order of the elements expressing the Cartesian coordinates $(i,j)$  in terms of a single label $p=i+(j-1)N$ that takes values $p=1,2, \dots N \times N$. With these notations, a second-order tensor 
can be represented as a vector 
and   a fourth-order tensor takes the form of  as  a  matrix. 

Consider for simplicity an externally unloaded body.   Using the lexicographic  notations we can  write  the  equilibrium  equation $\partial \Phi/\partial u=0$, in the form
$
K\zeta=h+\A(\xi-(d+g)),
$
where $\A$ is a standard forth order tensor with constant entries. Substituting the expression for $\zeta$  into the compatibility equations \eqref{stvenant} we obtain
$
 h+ \A(\xi-(d+g))=K\B \xi,
$
  where   $\B$ is another standard  forth order tensor;
the explicit expressions for the tensors $\A$ and $\B$  can be found   in \cite{Perez-Reche2016-xp}.

Using the obtained relations we  can  write the explicit representation of elastic solution in the form 
\begin{equation}
\xi=(\Id-K\A^{-1}\B )^{-1}\left[(d+g)-\A^{-1} h\right]=0.
\label{equil1}
\end{equation}
We  stress    that  the 'local' disorder $g$ 
enters \eqref{equil1}
as a quenched analog of a  compatible plastic deformation. Instead the incompatible disorder $h$ enters the solution nonlocally in the sense that 
 a residual  stress   $h$  placed  in the element  $\lbrace k,l\rbrace$  affects the actual elastic strain field $\xi$ in every other element  $\lbrace i,j \rbrace$. 
 
\begin{figure}[!htbp]
\centering
\includegraphics[scale=0.187]{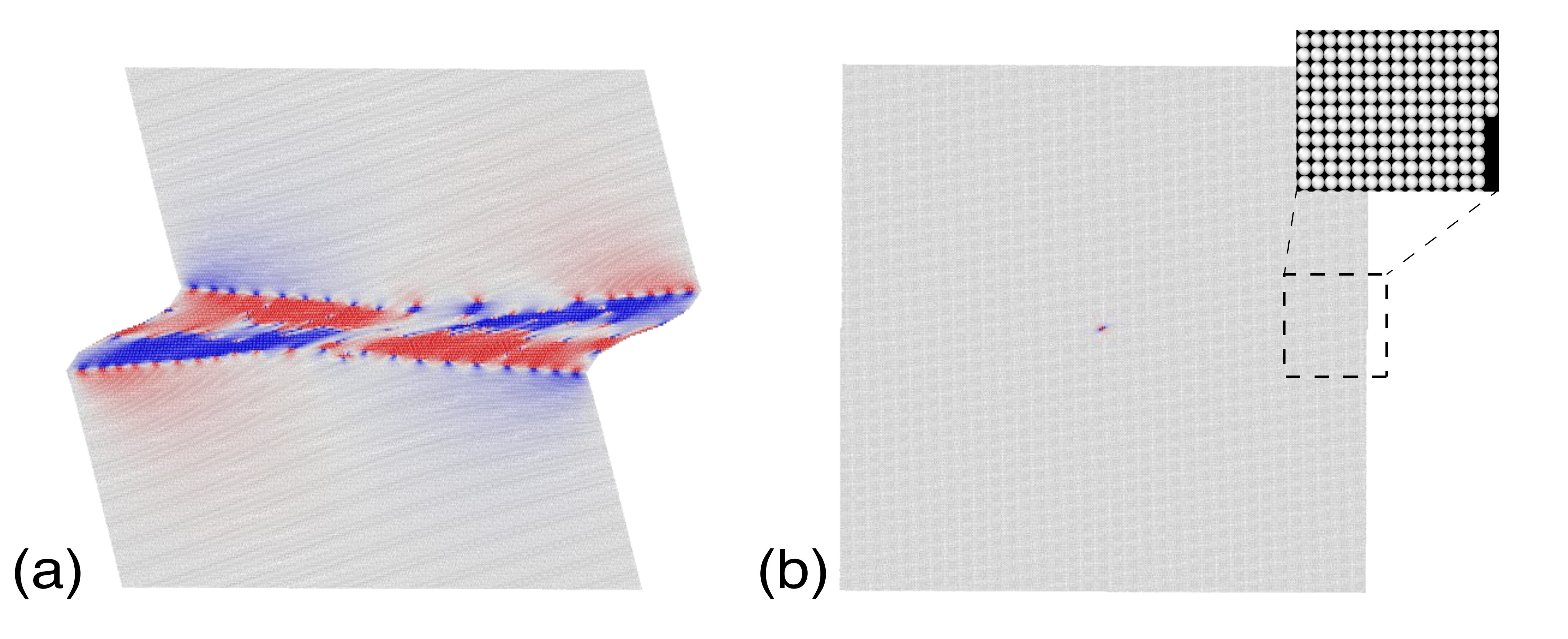}
\caption{Comparison of the displacement fields resulting from 
pointwise 'nonlocal' and 'local' disorders: (a) $g_{i,j} \equiv 0$ and $h_{i,j} = 0.5\delta_{i,0}\delta_{j,0}$;  (b) $g_{i,j} = 0.5\delta_{i,0}\delta_{j,0}$ and $h_{i,j} \equiv 0$. The 'nonlocal' disorder produces a  shear band with many dislocations. The 'local' disorder produces a  single slip due to  nucleation and subsequent annihilation of a pair of dislocations. Here $K=2$, $N=256$; colors reflect the level of longitudinal strain.
}
\label{kernel11}
\end{figure}

 Note also that since plastic slip develops  to minimize elastic energy, it  effectively acts to bring the expression in square brackets in \eqref{equil1} closer to zero. It can then compensate a compact  source of the disorder $g$ by yielding locally, at the location of such source. Instead, a compact source of the disorder $h$ can be compensated only by a broadly distributed plastic slip. To illustrate this point, we compare in Fig. \ref{kernel11}   the   responses  of a  loaded  crystal with  either    'nonlocal'  disorder $h$ or  'local'    disorder $g$  present   in the form of a point source. 

Specifically, we consider the disorder fields  $g_{i,j} = p\delta_{i,0}\delta_{j,0}$ and  $h_{i,j} = q\delta_{i,0}\delta_{j,0}$,  where $\delta_{i,j}$ is the Kronecker delta, and choose the  amplitudes  $p=q$  to ensure that when only one of these fields is present at $\gamma=0$  no plasticity occurs. Then, in each of these two cases, we find the  smallest  increment $ \delta\gamma$   initiating  a slip in  at least one  element. 

If  $p=0 $ and  $q=0.5$ (minimal  'nonlocal' disorder),  the avalanche resulting from such loading   is dramatic with many dislocations forming collectively
and the system developing a  \emph{macroscopic}  shear band with complex internal structure, see  Fig.~\ref{kernel11}(a).
If $p=0.5$ and  $q=0$  (minimal  'local' disorder),  two dislocations of opposite sign nucleate at the source  of the disorder  and move apart  to finally  annihilate  at the boundaries where we impose periodic boundary conditions.
Therefore the  response  remains contained and reduces to the formation of a \emph{microscopic} slip at the scale of a single element, see  Fig. \ref{kernel11}(b).

The observed nonlocal (global)  accommodation of the    disorder $h$ is possible only when the system is sufficiently homogeneous. In the presence of a substantial   'local' disorder $g$,  the ability of the system to generate such global  response  may  be compromised.  At sufficient strength of 'local' disorder  $\delta_g$ the coherent accommodation of 'nonlocal' disorder $\delta_h$  will   become  impossible, and as we argue below,  this can change  the avalanche scaling in a fundamental way.

\begin{figure}[!htbp]
	\centering	
	\includegraphics[scale=.65]{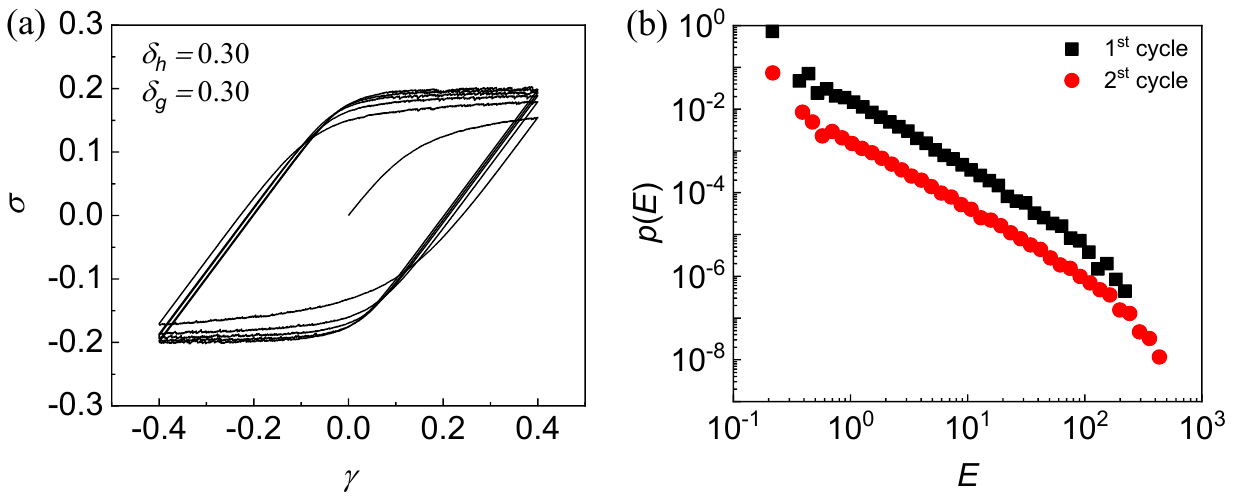}	
	\caption{(a) Strain-stress curves for the crystals subjected to six loading unloading cycles; (b) Avalanche distributions of cyclically loaded crystals for the first and the second cycles;  the first cycle is understood as  the monotone loading path.  Here $\delta_h=\delta=0.30$, $\delta_g=0.30$.
		\label{cyclic_distribution2}}
\end{figure}

To avoid any dependence on the initial preparation of the sample, we have chosen to present the interplay between the two types of disorder, 'local' and 'nonlocal'  in the setting of cyclic loading. Our numerical experiments, summarized in Fig. \ref{cyclic_distribution2}(a), show  that when a weak 'local' disorder $\delta_g=0.3$ is combined with a weak 'nonlocal'  disorder $\delta_h=0.3$,  the overall mechanical response is  ductile. The  initial softening behavior, observed in crystals with $\delta_g=0$,  is replaced by the more conventional  hardening behavior. At large strains   the stress response  shows  a robust yielding plateau independently of the configuration of disorder. The overall response  is  reminiscent of the  classical   strain-hardening behavior  exhibited   by \emph{bulk} FCC and BCC materials \cite{suresh1998fatigue}.    

From  Fig.~\ref{cyclic_distribution2}(b)  we see that even a weak 'local' disorder is  sufficient to  suppress  super-criticality and to completely eliminate system-size events.  This observation agrees with the idea that such disorder generates local inhomogeneities  which inhibit global  response. However, the increase of the cut-off size in the second cycle  suggests  that  a correlated behavior,  reminiscent of disorder-induced self-organization towards classical criticality  in RFIM model \cite{dahmen1996hysteresis,da2020rigidity}, can still take place.

\begin{figure}[!htbp]
\includegraphics[scale=0.26]{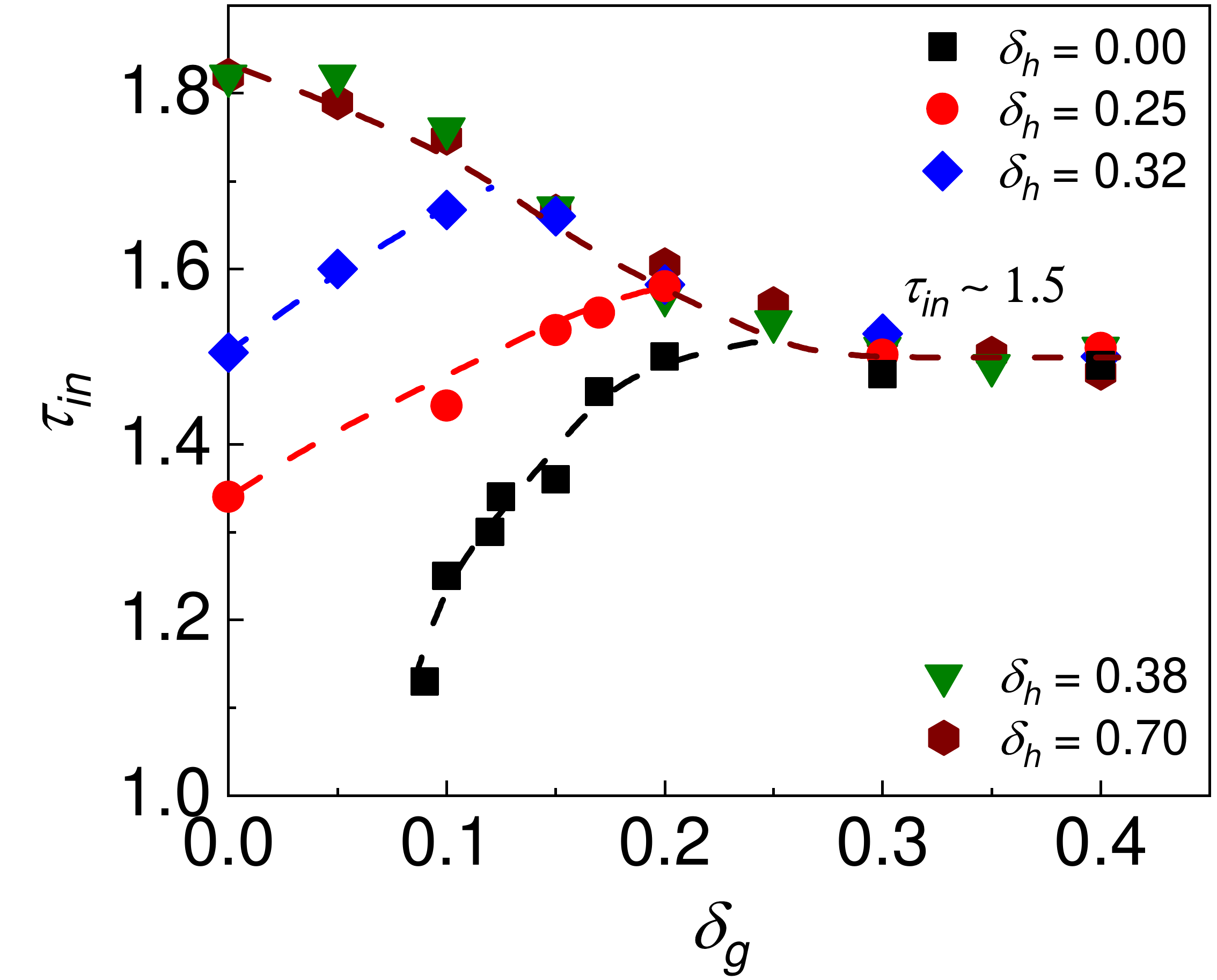}
\caption{   Effect of the 'local' disorder $\delta_g$ on the (integrated) scaling exponent $\tau_{in}$ for  the case of cyclic loading.
\label{fig811}   }
\end{figure}

In Fig. \ref{fig811} we show how the different  configurations of 'local' and  'nonlocal' disorder strengths  affect the cycle-averaged  (integrated) scaling exponents $\tau_{in}$. When the   'local' disorder is weak, we recover the after-yield behavior studied in Section \ref{CL}. At stronger  'local'  disorder, the dependence of the exponent $\tau_{in}$ on the  'nonlocal'  disorder progressively diminishes. Beyond  $\delta_g \sim  0.25$ it completely disappears, and the exponent  stabilizes around the value $\tau_{in} \sim 3/2$. Given that the statistics   is mostly  acquired during   hardening-free yield, see  Fig. \ref{cyclic_distribution2}, one can expect the stress-resolved value of the exponent $\tau$ to be similar to the aggregate value  $\tau_{in}$ \cite{durin2006role}.  In this case the  obtained exponent value  suggests  mean field criticality \cite{dahmen1996hysteresis,da2020rigidity}. In other words, the abundance  of 'local' disorder apparently   trivializes  the scaling picture, erasing the non-universality and promoting a universal response of the athermally driven infinite dimensional RFIM dominating the response of amorphous solids \cite{Ozawa2018-xi,ozawa2020role,bhaumik2019role,franz2020large}.

\section{Conclusions}
\label{C}

%
 
 
To  address the fundamental question why the dislocation avalanches  in sub-micron crystals of both face-centered cubic (fcc)  and  body-centered cubic (bcc) metals   exhibit 'wild' scaling, while  the associated bulk crystals are  'mild',  we conducted  a  range of numerical experiments using a minimal integer-automaton model of crystal plasticity.  

Our  approach to the study of the size effect is   based on the assumption  that the dominance of surface-induced dislocation activity in sub-micron crystals can be modeled by the scarcity of conventional bulk dislocation sources. To justify this assumption, we compared the effects of   extreme miniaturization in our physical experiments on Mo micro-pillars with the behavior of the numerical model as  we progressively  diminished the strength of quenched disorder. 
 In both cases, we observed the same second-order BD transition, which provides the basic explanation for the ultimate shift from ’wildness’ to ’mildness’ in the fluctuational response  with non-universality ultimately emerging as a size-effect. 


The  detailed transition from largely  brittle to mostly ductile  behavior  
was conceptualized  as a   complex three-stage crossover.   The individual transitions are from spin-glass-type marginality, characteristic of very small, almost  disorder free crystals,  through spinodal/depinning criticality at intermediate sizes (moderate disorder level),  to the classical BD  criticality in  larger,  highly disordered crystals. In general, this scenario  shows some similarity with the one observed  in  amorphous plasticity \cite{Ozawa2018-xi,Popovic2018-pp}, however, the  nuanced picture in crystal plasticity appears to be more intricate.

In  addition to monotone loading,  we also considered  large amplitude oscillatory shear loading protocols. We  observed that brittleness disappears after cyclic loading, which suggests that nominally  brittle  sub-micron crystals can be ’trained’ to become ductile. However, the basic crossover structure,  including the presence of three distinct universality classes, was shown  not to be affected by the type of loading. 




The  non-universality of the scaling behavior  progressively weakens  as  we complement  the generic incompatible  disorder, interacting with plasticity nonlocally,  by   the more special compatible  disorder, which affects the plastic slip locally as, for instance,  in RFIM model. Our numerical experiments  suggest that the  increase  in  the strength of the 'local' disorder   eventually restores  universality bringing the    system into the mean field  type criticality  class characteristic  of amorphous solids.

Our schematic scalar model would have to be extended to the full tensorial theory based on nonlinear elasticity to obtain a more detailed description of crystal plasticity. This extension will allow one to distinguish between different crystallographic classes and different configurations of dislocation cores. Such a model should be able to reproduce dislocation walls and generate cell structures with realistic size distribution. A 3D theory of this type can also capture dislocation  cross-slip and address hardening behavior.  To study  surface effects directly, one  would need to come up with the  boundary conditions allowing  for dislocation nucleation on the surface. A path in this general direction has been recently sketched in  \cite{Baggio2019-rs}.



%


\section{Acknowledgments} \label{A}

We thank N. Gorbushin, M. Mungan, F.  Perez Reche  and D.  Vandembroucq  for helpful discussions and Jin-yu Zhang for the characterization of dislocation structure. This work was supported by  
 French-Chinese ANR-NSFC grant (ANR-19-CE08-0010-01 and  51761135031).
  P.Z.  acknowledges additional support from China Scholarship Council and China Postdoctoral Science Foundation (grant 2019M653595). 

%


\end{document}